\RequirePackage{lineno}
\documentclass[twocolumn,aps,prd,superscriptaddress,showpacs]{revtex4}
\usepackage{epsfig}
\usepackage{graphicx}
\usepackage{dcolumn}
\usepackage{amsmath}
\usepackage{subfigure}
\usepackage{xspace}
\usepackage{color}
\usepackage{amssymb}
\usepackage{epsfig}
\usepackage{footnote}
\usepackage{longtable}
\usepackage{fancyhdr}
\usepackage{subfigure}
\usepackage{xspace}
\usepackage{multirow}
\usepackage{comment}

\newcommand{\gev}{\ensuremath{\mbox{GeV}}}

\newcommand{\pgev}{\ensuremath{\mbox{GeV/}\mbox{c}}}
\newcommand{\mgev}{\ensuremath{\mbox{GeV}/\mbox{c}}^{2}}

\newcommand{\iet}{\ensuremath{E_{T}}}

\newcommand{\ippt}{p_{T}}

\newcommand{\qqbar}{\ensuremath{q\overline{q}}}
\newcommand{\ccbar}{\ensuremath{c\overline{c}}}
\newcommand{\bbbar}{\ensuremath{b\overline{b}}}
\newcommand{\ppbar}{\ensuremath{p\overline{p}}}
\newcommand{\ttbar}{\ensuremath{t\overline{t}}}

\begin{document}


\title{Search for heavy metastable particles decaying to jet pairs in \ppbar\ collisions at $\sqrt{s}=1.96\ \mbox{TeV}$}

\affiliation{Institute of Physics, Academia Sinica, Taipei, Taiwan 11529, Republic of China} 
\affiliation{Argonne National Laboratory, Argonne, Illinois 60439, USA} 
\affiliation{University of Athens, 157 71 Athens, Greece} 
\affiliation{Institut de Fisica d'Altes Energies, ICREA, Universitat Autonoma de Barcelona, E-08193, Bellaterra (Barcelona), Spain} 
\affiliation{Baylor University, Waco, Texas 76798, USA} 
\affiliation{Istituto Nazionale di Fisica Nucleare Bologna, $^z$University of Bologna, I-40127 Bologna, Italy} 
\affiliation{University of California, Davis, Davis, California 95616, USA} 
\affiliation{University of California, Los Angeles, Los Angeles, California 90024, USA} 
\affiliation{Instituto de Fisica de Cantabria, CSIC-University of Cantabria, 39005 Santander, Spain} 
\affiliation{Carnegie Mellon University, Pittsburgh, Pennsylvania 15213, USA} 
\affiliation{Enrico Fermi Institute, University of Chicago, Chicago, Illinois 60637, USA}
\affiliation{Comenius University, 842 48 Bratislava, Slovakia; Institute of Experimental Physics, 040 01 Kosice, Slovakia} 
\affiliation{Joint Institute for Nuclear Research, RU-141980 Dubna, Russia} 
\affiliation{Duke University, Durham, North Carolina 27708, USA} 
\affiliation{Fermi National Accelerator Laboratory, Batavia, Illinois 60510, USA} 
\affiliation{University of Florida, Gainesville, Florida 32611, USA} 
\affiliation{Laboratori Nazionali di Frascati, Istituto Nazionale di Fisica Nucleare, I-00044 Frascati, Italy} 
\affiliation{University of Geneva, CH-1211 Geneva 4, Switzerland} 
\affiliation{Glasgow University, Glasgow G12 8QQ, United Kingdom} 
\affiliation{Harvard University, Cambridge, Massachusetts 02138, USA} 
\affiliation{Division of High Energy Physics, Department of Physics, University of Helsinki and Helsinki Institute of Physics, FIN-00014, Helsinki, Finland} 
\affiliation{University of Illinois, Urbana, Illinois 61801, USA} 
\affiliation{The Johns Hopkins University, Baltimore, Maryland 21218, USA} 
\affiliation{Institut f\"{u}r Experimentelle Kernphysik, Karlsruhe Institute of Technology, D-76131 Karlsruhe, Germany} 
\affiliation{Center for High Energy Physics: Kyungpook National University, Daegu 702-701, Korea; Seoul National University, Seoul 151-742, Korea; Sungkyunkwan University, Suwon 440-746, Korea; Korea Institute of Science and Technology Information, Daejeon 305-806, Korea; Chonnam National University, Gwangju 500-757, Korea; Chonbuk National University, Jeonju 561-756, Korea} 
\affiliation{Ernest Orlando Lawrence Berkeley National Laboratory, Berkeley, California 94720, USA} 
\affiliation{University of Liverpool, Liverpool L69 7ZE, United Kingdom} 
\affiliation{University College London, London WC1E 6BT, United Kingdom} 
\affiliation{Centro de Investigaciones Energeticas Medioambientales y Tecnologicas, E-28040 Madrid, Spain} 
\affiliation{Massachusetts Institute of Technology, Cambridge, Massachusetts 02139, USA} 
\affiliation{Institute of Particle Physics: McGill University, Montr\'{e}al, Qu\'{e}bec, Canada H3A~2T8; Simon Fraser University, Burnaby, British Columbia, Canada V5A~1S6; University of Toronto, Toronto, Ontario, Canada M5S~1A7; and TRIUMF, Vancouver, British Columbia, Canada V6T~2A3} 
\affiliation{University of Michigan, Ann Arbor, Michigan 48109, USA} 
\affiliation{Michigan State University, East Lansing, Michigan 48824, USA}
\affiliation{Institution for Theoretical and Experimental Physics, ITEP, Moscow 117259, Russia}
\affiliation{University of New Mexico, Albuquerque, New Mexico 87131, USA} 
\affiliation{Northwestern University, Evanston, Illinois 60208, USA} 
\affiliation{The Ohio State University, Columbus, Ohio 43210, USA} 
\affiliation{Okayama University, Okayama 700-8530, Japan} 
\affiliation{Osaka City University, Osaka 588, Japan} 
\affiliation{University of Oxford, Oxford OX1 3RH, United Kingdom} 
\affiliation{Istituto Nazionale di Fisica Nucleare, Sezione di Padova-Trento, $^{aa}$University of Padova, I-35131 Padova, Italy} 
\affiliation{LPNHE, Universite Pierre et Marie Curie/IN2P3-CNRS, UMR7585, Paris, F-75252 France} 
\affiliation{University of Pennsylvania, Philadelphia, Pennsylvania 19104, USA}
\affiliation{Istituto Nazionale di Fisica Nucleare Pisa, $^{bb}$University of Pisa, $^{cc}$University of Siena and $^{dd}$Scuola Normale Superiore, I-56127 Pisa, Italy} 
\affiliation{University of Pittsburgh, Pittsburgh, Pennsylvania 15260, USA} 
\affiliation{Purdue University, West Lafayette, Indiana 47907, USA} 
\affiliation{University of Rochester, Rochester, New York 14627, USA} 
\affiliation{The Rockefeller University, New York, New York 10065, USA} 
\affiliation{Istituto Nazionale di Fisica Nucleare, Sezione di Roma 1, $^{ee}$Sapienza Universit\`{a} di Roma, I-00185 Roma, Italy} 

\affiliation{Rutgers University, Piscataway, New Jersey 08855, USA} 
\affiliation{Texas A\&M University, College Station, Texas 77843, USA} 
\affiliation{Istituto Nazionale di Fisica Nucleare Trieste/Udine, I-34100 Trieste, $^{ff}$University of Trieste/Udine, I-33100 Udine, Italy} 
\affiliation{University of Tsukuba, Tsukuba, Ibaraki 305, Japan} 
\affiliation{Tufts University, Medford, Massachusetts 02155, USA} 
\affiliation{University of Virginia, Charlottesville, Virginia  22906, USA}
\affiliation{Waseda University, Tokyo 169, Japan} 
\affiliation{Wayne State University, Detroit, Michigan 48201, USA} 
\affiliation{University of Wisconsin, Madison, Wisconsin 53706, USA} 
\affiliation{Yale University, New Haven, Connecticut 06520, USA} 
\author{T.~Aaltonen}
\affiliation{Division of High Energy Physics, Department of Physics, University of Helsinki and Helsinki Institute of Physics, FIN-00014, Helsinki, Finland}
\author{B.~\'{A}lvarez~Gonz\'{a}lez$^v$}
\affiliation{Instituto de Fisica de Cantabria, CSIC-University of Cantabria, 39005 Santander, Spain}
\author{S.~Amerio}
\affiliation{Istituto Nazionale di Fisica Nucleare, Sezione di Padova-Trento, $^{aa}$University of Padova, I-35131 Padova, Italy} 

\author{D.~Amidei}
\affiliation{University of Michigan, Ann Arbor, Michigan 48109, USA}
\author{A.~Anastassov}
\affiliation{Northwestern University, Evanston, Illinois 60208, USA}
\author{A.~Annovi}
\affiliation{Laboratori Nazionali di Frascati, Istituto Nazionale di Fisica Nucleare, I-00044 Frascati, Italy}
\author{J.~Antos}
\affiliation{Comenius University, 842 48 Bratislava, Slovakia; Institute of Experimental Physics, 040 01 Kosice, Slovakia}
\author{G.~Apollinari}
\affiliation{Fermi National Accelerator Laboratory, Batavia, Illinois 60510, USA}
\author{J.A.~Appel}
\affiliation{Fermi National Accelerator Laboratory, Batavia, Illinois 60510, USA}
\author{A.~Apresyan}
\affiliation{Purdue University, West Lafayette, Indiana 47907, USA}
% ...
\collaboration{CDF Collaboration\footnote{With visitors from $^a$University of Massachusetts Amherst, Amherst, MA 01003, USA.
$^b$Istituto Nazionale di Fisica Nucleare, Sezione di Cagliari, 09042 Monserrato (Cagliari), Italy.
$^c$University of California Irvine, Irvine, CA  92697, USA.
$^d$University of California Santa Barbara, Santa Barbara, CA 93106, USA.
$^e$University of California Santa Cruz, Santa Cruz, CA  95064, USA.
$^f$CERN,CH-1211 Geneva, Switzerland.
$^g$Cornell University, Ithaca, NY  14853, USA.
$^h$University of Cyprus, Nicosia CY-1678, Cyprus.
$^i$University College Dublin, Dublin 4, Ireland.
$^j$University of Fukui, Fukui City, Fukui Prefecture, Japan 910-0017.
$^k$Universidad Iberoamericana, Mexico D.F., Mexico.
$^l$Iowa State University, Ames, IA  50011, USA.
$^m$University of Iowa, Iowa City, IA  52242, USA.
$^n$Kinki University, Higashi-Osaka City, Japan 577-8502.
$^o$Kansas State University, Manhattan, KS 66506, USA.
$^p$University of Manchester, Manchester M13 9PL, England.
$^q$Queen Mary, University of London, London, E1 4NS, England.
$^r$Muons, Inc., Batavia, IL 60510, USA.
$^s$Nagasaki Institute of Applied Science, Nagasaki, Japan.
$^t$National Research Nuclear University, Moscow, Russia.
$^u$University of Notre Dame, Notre Dame, IN 46556, USA.
$^v$Universidad de Oviedo, E-33007 Oviedo, Spain.
$^w$Texas Tech University, Lubbock, TX  79609, USA.
$^x$Universidad Tecnica Federico Santa Maria, 110v Valparaiso, Chile.
$^y$Yarmouk University, Irbid 211-63, Jordan.
$^{gg}$On leave from J.~Stefan Institute, Ljubljana, Slovenia.
}}
\noaffiliation

\date{\today}

\begin{abstract}
A search is performed for heavy metastable particles that decay
into jet pairs with a macroscopic lifetime ($c\tau\sim 1\ \mbox{cm}$)
in \ppbar\ collisions at $\sqrt{s}=1.96\ \mbox{TeV}$ using data from
the CDF~II detector at Fermilab corresponding to an integrated
luminosity of $3.2\ \mbox{fb}^{-1}$. To estimate the standard model
background a data-driven approach is used. Probability-density
functions are constructed to model secondary vertices from known
processes. No statistically significant excess is observed above the
background. Limits on the production cross section in a hidden valley
benchmark phenomenology are set for various Higgs boson masses as well
as metastable particle masses and lifetimes.
\end{abstract}

\maketitle


\section{Introduction}
\label{sect:introduction}

The standard model (SM) of elementary particles fails at TeV energies
if no new phenomena appear at this scale. A single Higgs boson
provides the simplest solution, but there are other possibilities,
some of which predict massive metastable particles. They are
metastable because they can only decay to SM particles through
diagrams containing a new high-mass force carrier or a loop of very
massive particles. These model are broadly categorized as ``hidden
valley'' (HV) models~\cite{Strassler2007374}. We use data from \ppbar\
collisions at $\sqrt{s}=1.96\ \mbox{TeV}$ collected with the CDF~II
detector at the Fermilab Tevatron collider to search for a long-lived
massive particle that originates from the primary \ppbar\ interaction,
travels a macroscopic distance (of order $1\ \mbox{cm}$), and decays
into jet pairs. A variety of predicted decay modes are possible for
these metastable particles. Although the search is sensitive to any
massive long-lived particle decaying into jet pairs, for the sake of
specificity we choose as a benchmark to evaluate the results within
the context of the HV phenomenology.

A recent analysis from the D0 experiment searched for heavy particles
decaying with a displaced vertex~\cite{PhysRevLett.103.071801}, using
the same phenomenological model as this analysis.  However it was
restricted to heavy metastable particles that decay into $b$ quarks
because their trigger required a muon in the event.  We have no such
limitation because CDF employs the silicon vertex trigger (SVT) which
allows us to trigger on tracks that originate from displaced
vertices~\cite{Bardi:1997hm, Adelman:2007zz}. Thus our search is sensitive to
metastable particles that decay into any jets, not only $b$-quark
jets.

We search for an event signature where two jets of particles emanate
from a point displaced from the primary interaction point, i.e., a
displaced or secondary vertex. Monte Carlo (MC) simulation is used to
model events from the HV phenomenology. It serves as our benchmark for
processes containing the signature of the search. Since the SM does
not contain massive metastable particles, we expect little
background. We construct a background model almost entirely from the
data. To accomplish this, the kinematics of the data events with
secondary vertex characteristics are determined from auxiliary SM
samples.

After a brief overview of the CDF~II detector in
Sec.~\ref{sect:detector}, the HV phenomenology is described in
Sec.~\ref{sect:hiddenvalley}.  Section~\ref{sect:eventselection}
discusses the event selection for the analysis, and
Sec.~\ref{sect:backgroundestimation} presents the background
estimate along with a test of the method.  The search for the signal
is presented in Sec.~\ref{sect:signalsearch}.  Systematic
uncertainties for both the expected signal and the background are
presented in
Sec.~\ref{sect:systematicuncertainties}. Section~\ref{sect:conclusion}
presents limits on the production cross-section in the HV
phenomenology.

\section{The CDF Detector}
\label{sect:detector}

CDF is a general-purpose detector that is described in detail in
Ref.~\cite{CDF}. The detector components relevant to this analysis are
briefly described here. Closest to the beampipe are multilayer silicon
detectors (SVX)~\cite{SVX} providing precision tracking which is used
to identify displaced vertices. Outside the multilayer silicon
detectors is an open-cell drift chamber, the central outer tracker
(COT), covering the pseudorapidity region $|\eta| <
1$~\cite{COT}. (The pseudorapidity $\eta$ is defined as
$-\ln[\tan(\theta/2)]$, where $\theta$ is the polar angle relative to
the proton beam direction~\cite{coord}.) The COT is used to
reconstruct charged particles' momenta. The tracking system is
enclosed in a superconducting solenoid operating at $1.4\ \mbox{T}$,
which in turn is surrounded by a calorimeter.

The CDF calorimeter system is separated into electromagnetic and
hadronic components segmented in a projective tower geometry
covering the region $|\eta| < 3.6$. The electromagnetic calorimeters
use a lead-scintillator sampling technology~\cite{CEM}, whereas
the hadron calorimeters use iron-scintillator
technology~\cite{CHA,cal_upgrade}. Jets are reconstructed from
the energy deposited in these calorimeters~\cite{jetclu}. The
calorimeter is separated into the central ($|\eta|<1.0$) and forward
(or plug) regions ($|\eta|>1.0$).

Finally, the muon subdetectors are arrayed outside the
calorimeters. The beam luminosity is determined with gas Cherenkov
counters located in the region $3.7 < |\eta| < 4.7$ which measure the
average number of inelastic \ppbar\ collisions per accelerator bunch
crossing~\cite{CLC}.

CDF uses a three-level trigger system, with a mix of hardware
electronics and dedicated CPUs to select interesting events. In our
analysis, a trigger sensitive to $Z \rightarrow b\bar{b}$ decays is
used, the ZBB trigger. It employs the SVT hardware in the
second-level trigger. This trigger is described in more detail in
Sec.~\ref{sect:eventselection}.




\section{The Hidden Valley Model}
\label{sect:hiddenvalley}

\subsection{Phenomenology} 
\label{subsec:HV_phenomenology}

While the analysis presented here is a search for any heavy particle
that decays into a pair of jets at a displaced vertex, it is useful to
have a benchmark model. The HV phenomenology provides a framework in
which we can generate signal Monte Carlo samples, search for
discriminants, optimize our search, and compare results. Results
presented for this benchmark process can be used to constrain other
models by accounting for the differences in the kinematic properties
of the final state. Here we present a brief outline of the HV picture.

In the HV scenario, the standard model gauge group $G_{SM}$ is
extended by a non Abelian group
$G_{v}$~\cite{Strassler2007374,Strassler2008263}. SM particles
are neutral under $G_{v}$. Additionally, $G_{v}$ contains new
particles, called $v$-particles, that are charged under $G_{v}$ but
neutral under $G_{SM}$. 

In the particular class of hidden valley models considered here, the
$G_{v}$ gauge group may become strong and confine, analogously to
QCD. The $v$-particles, called $v$-quarks in this class of models, are
confined inside $v$-hadrons. Energetic collisions at the Tevatron could
create new particles, such as the Higgs boson or a new $Z'$ resonance,
that could decay to HV particles. If the lightest $v$-hadrons are
sufficiently heavy, they can decay to SM particles via highly
suppressed processes, e.g., mixing with the longitudinal component of
a $Z'$. A wide range of masses, lifetimes, and final states are
possible within the HV framework. If the lightest available HV
particle, a $v$-$\pi$ in our benchmark model, has mass less than twice
that of the top quark, the predominant decay would be to \bbbar\ quark
pairs. With a long lifetime, some particles would travel a measurable
distance from the primary vertex before decaying, much like a $B$ or
$D$ hadron.

Applying the HV phenomenology to astroparticle physics, the authors in
Ref.~\cite{kaplanlutyzurek} theorize that the existence of a dark
matter candidate implies another (HV) particle with a lifetime such
that $c\tau$ is of the order of $1\ \mbox{cm}$. This places the
lifetime within the range accessible to the CDF~II detector.

\begin{figure}[htb]
\includegraphics[width=3.4in]{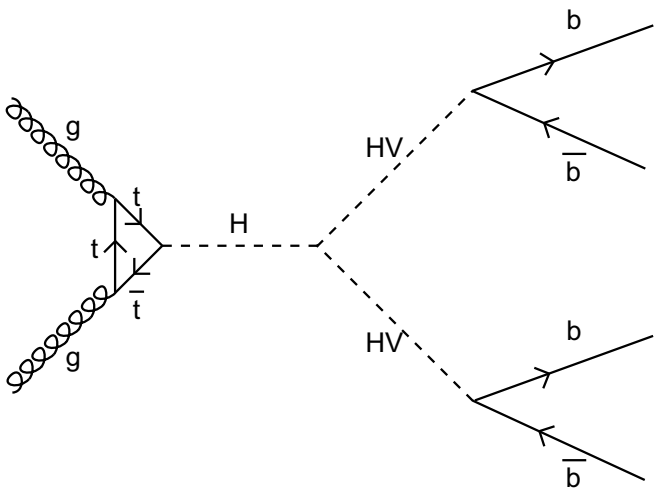}
\caption{
\label{fig_HVFeynman} Feynman diagram of Higgs boson production of a hidden valley particle and its subsequent decay. The coupling of HV particle to \bbbar\ is extremely small, resulting in the long lifetime.}
\end{figure}

The HV also provides a way to search for the Higgs boson. If the Higgs
boson mixes with a scalar in the HV sector that couples to $v$-quarks,
then it may decay to two (or more) $v$-hadrons. These $v$-hadrons would in
turn decay into \bbbar\ quark pairs. It would be feasible to search
for the Higgs boson using this final state at CDF. Under some HV
scenarios, the branching fraction to HV particles could be comparable
to those of SM decays. In addition, searches for these HV decays may
have higher signal to background ratios due to their unique decay
topology. A Feynman diagram of this decay is shown in
Fig.~\ref{fig_HVFeynman}.

\subsection{Monte Carlo Simulation Samples} 
\label{subsec:HV_MCsamples}

The {\sc Pythia} Monte Carlo program version 6.2~\cite{pythia} is
used to generate the events for the signal MC simulation. {\sc Geant3}
is used for the detector simulation~\cite{GEANT3}.  To mimic HV
production and decay in {\sc Pythia}, we use the minimal
supersymmetric standard model process of a {\it CP}-even Higgs boson
($h^{0}$) decaying into two {\it CP}-odd Higgs bosons ($a^{0}$) which in
turn decay into $b$-quarks, $h^{0} \rightarrow a^{0}a^{0} \rightarrow
b \bar{b}\ b \bar{b}$. We alter the mass and lifetime of the $a^{0}$
in order to simulate the HV particle. This allows for the generation
of signal MC samples using the {\sc Pythia} MC generator without
significant modification.

Two Higgs boson masses are generated, one at relatively low mass,
$130\ \mgev$, and one at higher mass, $170\ \mgev$. Multiple HV
particle masses from $20\ \mgev$ to $65\ \mgev$\ are produced. The
$c\tau$ of the HV particle, $c\tau_{\mathrm{HV}}$, is set to $1.0\
\mbox{cm}$. Some signal MC samples have been weighted to study HV
particles with $c\tau_{\mathrm{HV}}$ of $0.3$, $2.5$, or $5.0\ \mbox{cm}$. Thus
we can study multiple HV lifetimes without generating additional
signal MC samples.

\subsection{Discriminants from signal MC simulation} 
\label{subsec:HV_discriminants}

\begin{figure*}[htb]
\includegraphics[width=\textwidth]{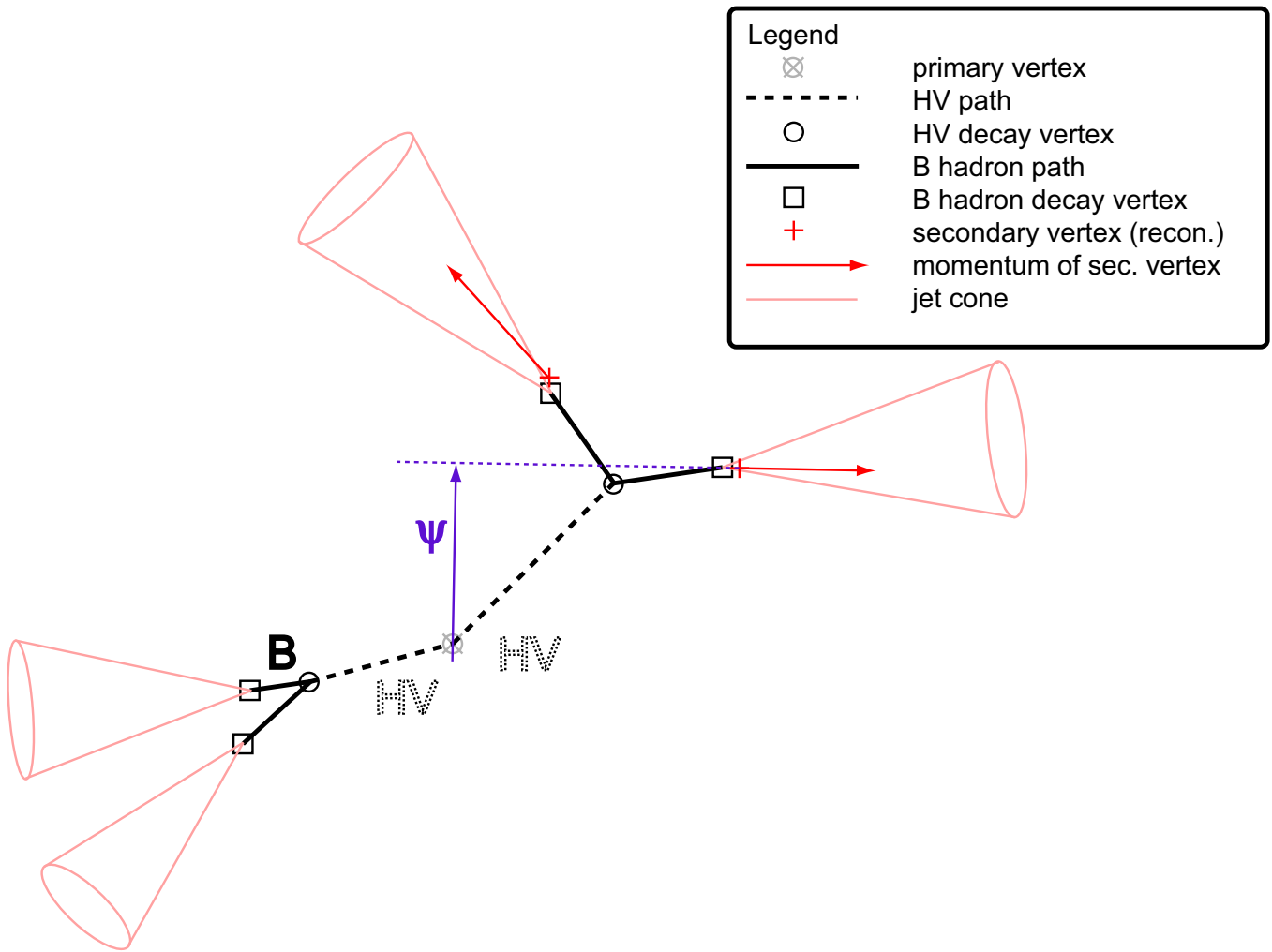}
\caption{
\label{fig_HVDisplacedVertexA06L} Schematic diagram of variable $\psi$, 
the impact parameter of a jet with a secondary vertex. This figure is
not to scale. The figure is shown in a plane transverse to the
beam line.}
\end{figure*}

\begin{figure*}[htb]
\includegraphics[width=\textwidth]{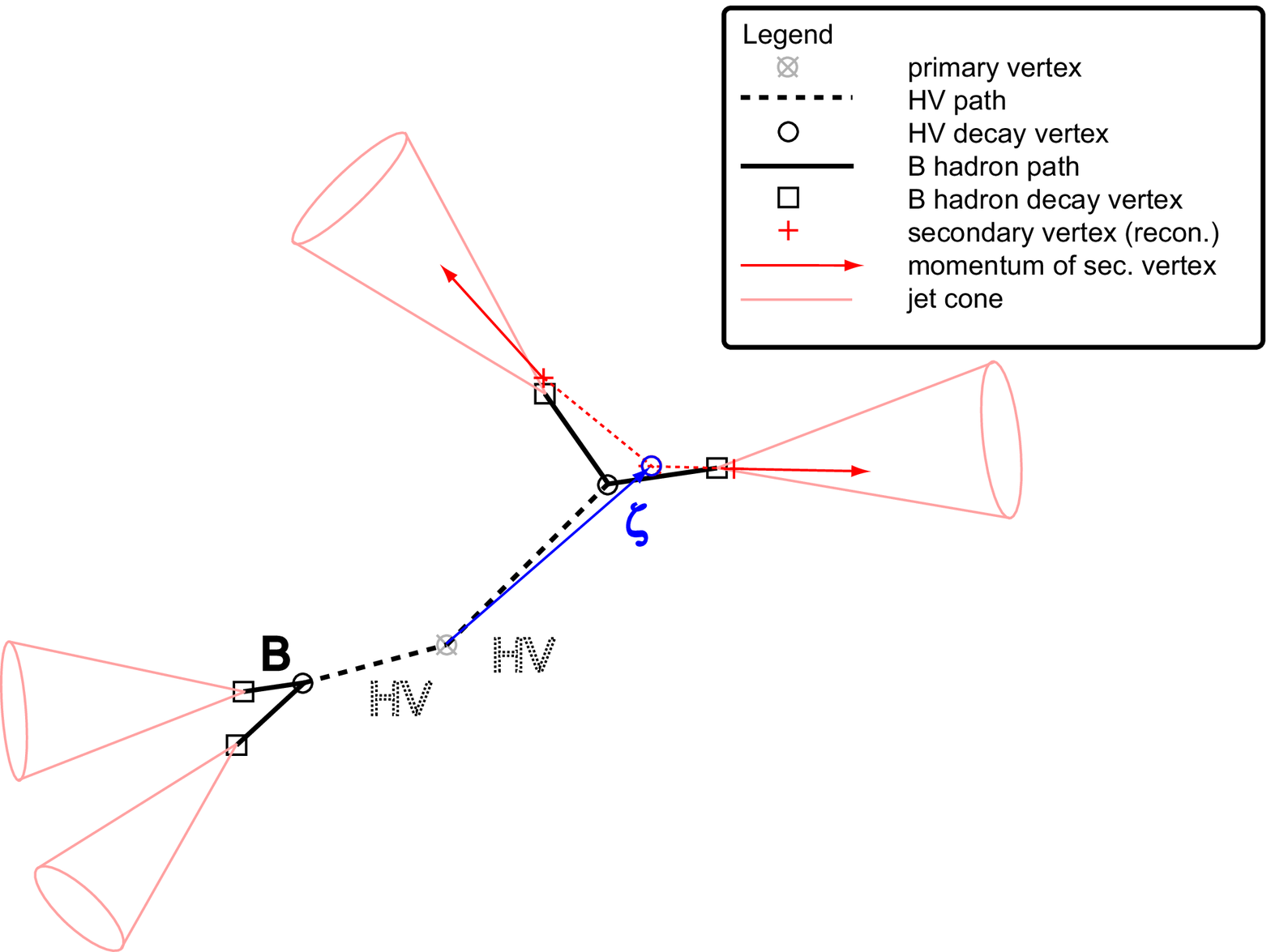}
\caption{
\label{fig_HVDisplacedVertexA03L} Schematic diagram of variable 
$\vec{\zeta}$, which represents the decay vertex of the HV particle.}
\end{figure*}

The major characteristic that distinguishes the signal from the SM
backgrounds is the presence of two jets whose momentum vectors both
point to a common secondary vertex. With this in mind, we developed two
discriminants shown in Fig.~\ref{fig_HVDisplacedVertexA06L}
and Figs.~\ref{fig_HVDisplacedVertexA03L}: $\psi$ and $\zeta$. In both
cases, the figures are drawn in the two-dimensional plane transverse
to the beam line.

We define a ``tagged jet'' as a jet with a reconstructed secondary vertex
using criteria defined in Sec.~\ref{sect:eventselection}. By
definition, such a jet has both a position (the secondary vertex) and a
direction defined by the sum of momenta of the tracks that make up the
vertex, where all these quantities are defined in the transverse plane.
We define $\vec{\psi}$ as the orthogonal vector from the primary vertex,
the reconstructed location of the \ppbar\ collision, to the line defined
by the secondary vertex position and momentum direction. The magnitude of
$\vec{\psi}$ is the distance from the primary vertex to the line, i.e.,
its impact parameter. We take the sign of $\psi$ as that of the dot
product between $\vec{\psi}$ and the momentum of the tagged jet. The
distribution of $\psi$ for simulated signal events has much larger tails
than simulated background events.

The variable $\zeta$ is defined for events where there are two tagged
jets; see Fig.~\ref{fig_HVDisplacedVertexA03L}. The intersection of
the two tag momenta can be thought of as the reconstructed decay
vertex of the HV particle. The vector from the primary vertex to this
reconstructed decay vertex is $\vec{\zeta}$; the magnitude, 
$\zeta$, is the reconstructed two-dimensional decay distance of
the HV particle.

The sign of $\zeta$ is determined by taking the dot product between
$\vec{\zeta}$ and the vector sum of the momenta of the two jets with
tags. The sign effectively indicates whether or not the decay vertex
is in the same hemisphere of the detector as the jet pair. Signal MC
events have more positive $\zeta$ than negative, while the background
MC events have $\zeta$ uniformly distributed around zero.

At this point, it is necessary to discuss the combinatorics of the HV
event topology. With MC simulation, we can use the generator-level
information to evaluate if the jets with secondary vertices originated
from quarks whose mother is the HV particle. Using this information, we
define four possible topologies in which signal MC events can be
classified.

\newcounter{SMCTOPOLOGY2}
\begin{list}{\arabic{SMCTOPOLOGY2}.}{\usecounter{SMCTOPOLOGY2}}
\item Two-tag HV: both jets originate from the {\bf same} HV particle.
\item One tag each: each jet originates from a different HV particle.
\item One HV jet: one jet originates from a HV particle; the other does not.
\item No HV jets: neither jet originates from a HV particle.
\end{list}

\begin{figure}[!htb]
\includegraphics[width=3.4in]{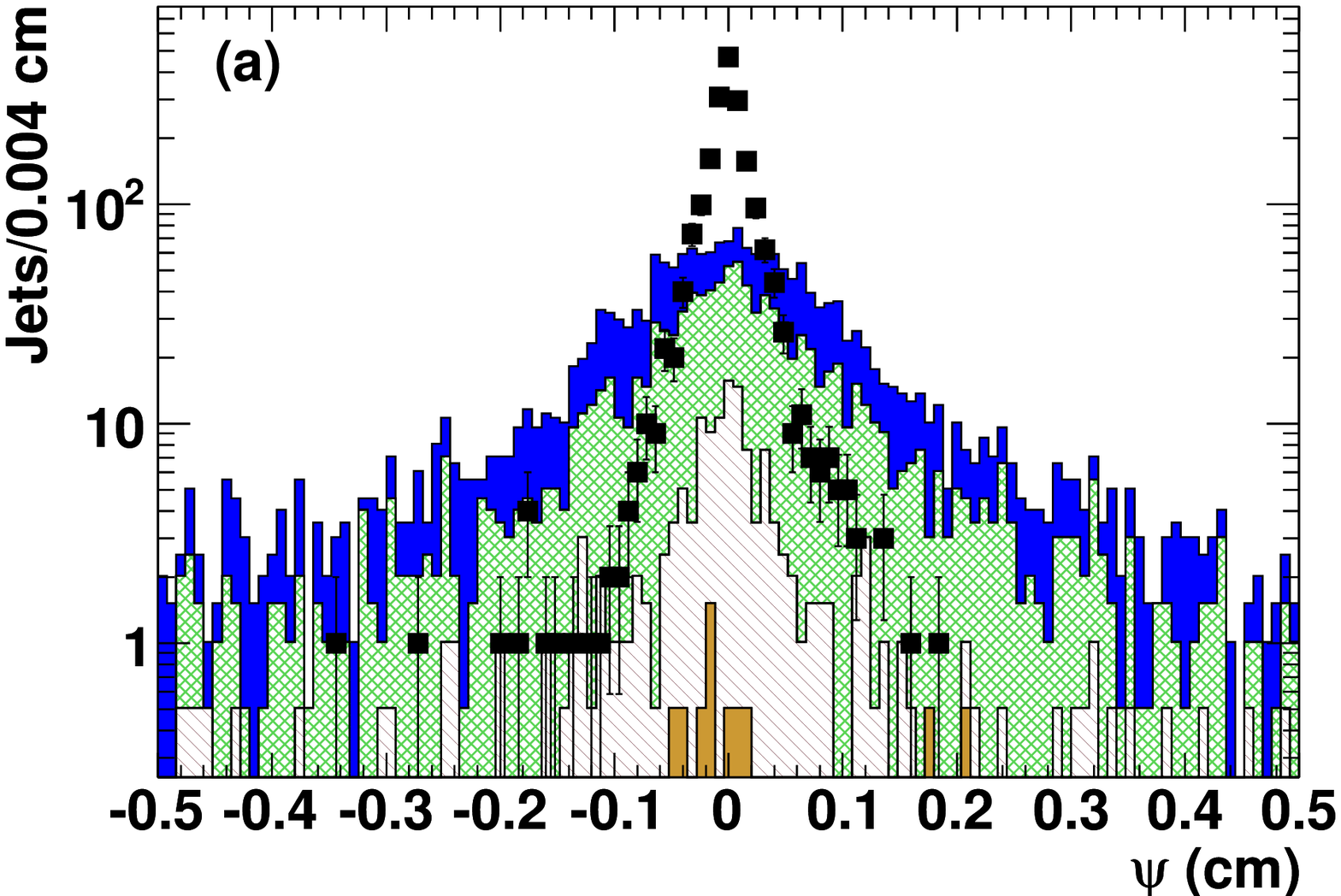}
\includegraphics[width=3.4in]{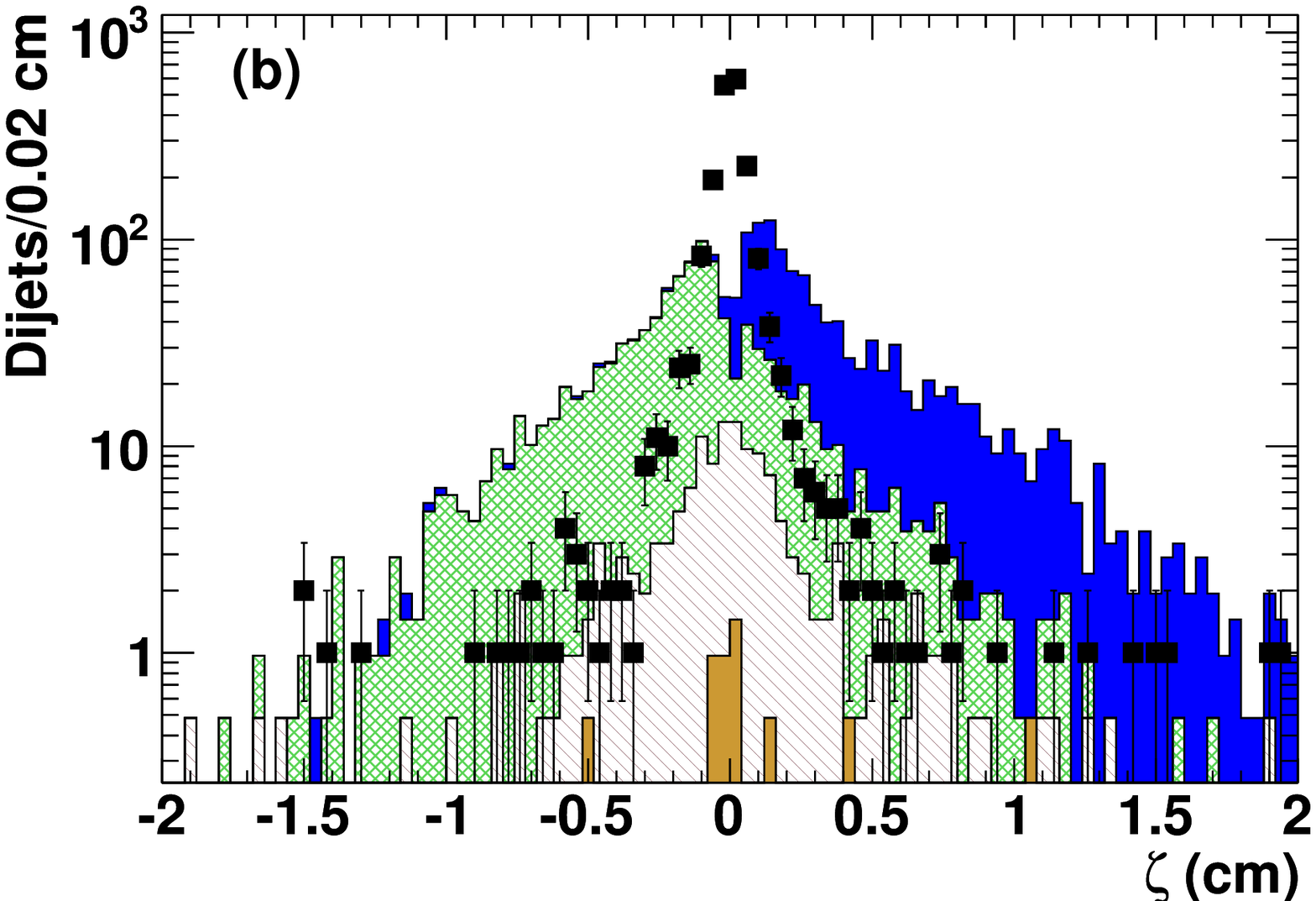}
\includegraphics[width=3.4in]{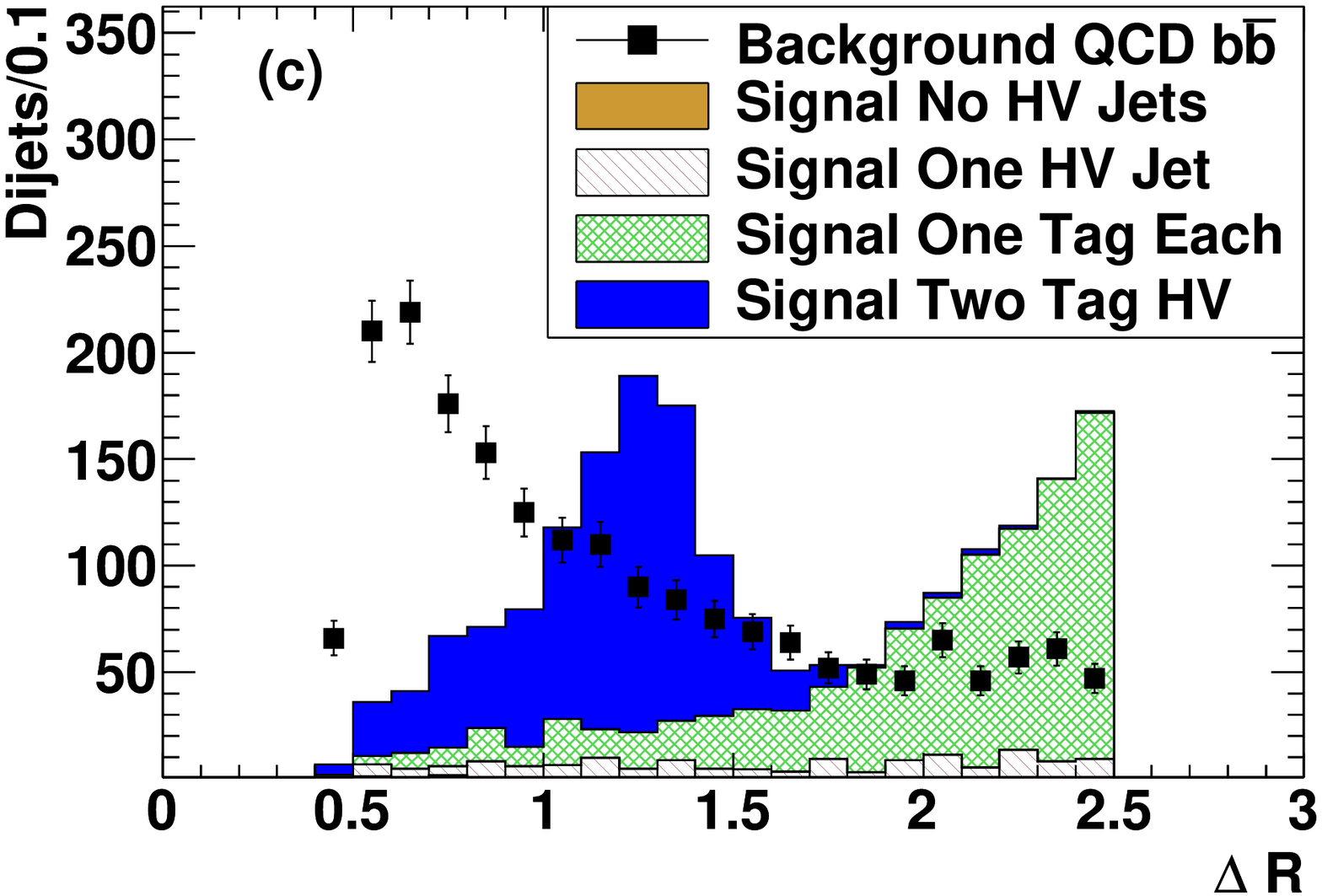}
\caption{ 
\label{fig_HVdists} 
Distributions of (a) $\psi$, (b) $\zeta$, and (c) $\Delta R$ for dijets
where both jets are tagged. The signal MC simulation histograms are
stacked. In (a), only the leading jet of the jet pair is shown. In
(c), a $\Delta R < 2.5$ cut has already been applied.  }
\end{figure}

\noindent Figure~\ref{fig_HVdists} shows the (a) $\psi$ and (b)
$\zeta$ distributions for different signal MC simulation topologies
and background MC simulation ({\sc Pythia} QCD \bbbar). The signal MC
assumes $M_{H}=130\ \mgev$, $M_{\mathrm{HV}}=40\ \mgev$, and
$c\tau_{\mathrm{HV}}=1.0\ \mbox{cm}$. These distributions have been normalized
to unit area and show the discriminating power of both variables.

For the two-tag HV topology, the distribution of $\zeta$ is nearly
always positive, which improves the discrimination against the
background. The one-tag-each topology is a distribution that is both
positive and negative, but mostly negative. Between the two-tag HV and
the one-tag-each topologies, we concern ourselves with the first
because it has more discriminating power, and if a signal is seen, the
HV particle's invariant mass can be reconstructed. The final two
topologies are very rare, but are included in the figures for
completeness.

In addition to these two variables, the separation of the two jets
($\Delta R = \sqrt{(\Delta \phi)^{2} + (\Delta \eta)^{2}}$) is a
useful discriminant (see Fig.~\ref{fig_HVdists}c). In the
two-tag HV topology, the decay daughters of the HV particle are more
collinear than the one-tag-each topology, but still different than
the QCD background, which is dominated by gluon splitting at low
$\Delta R$, once a $\Delta R<2.5$ cut removes most of the direct
\bbbar\ production.

\section{Event Selection}
\label{sect:eventselection}

\subsection{ZBB trigger}
\label{subsec:ES_zbbtrigger}

$Z \rightarrow b\bar{b}$ events are collected at CDF for the purpose
of studying the jet energy scale (JES) of $b$-quark
jets~\cite{Donini2008nt}~\cite{Bhatti2005ai}. This is achieved by
means of a specially designed trigger, the ZBB trigger
(Table~\ref{table_evzbbtrigger}) which selects events containing
tracks with a large impact parameter with respect to the primary
vertex ($d_{0}$). Because a HV particle would decay at a displaced
vertex, tracks from this decay would have large $d_{0}$.  Thus we use
this trigger for our signal search. The total integrated luminosity
collected with the ZBB trigger is $3.2\ \mbox{fb}^{-1}$.

At level\_1, the trigger has two requirements. It selects events with
at least one central calorimeter tower with $\iet\ > 5\
\gev$ and at least two tracks, one with $\ippt\ > 5.48\
\pgev$, the other one with $\ippt\ > 2.46\ \pgev$.

\begin{table*}[htb]
\caption{\label{table_evzbbtrigger} 
ZBB trigger requirements. One of the two level\_2 paths, opposite side or same side must be satisfied.}
\begin{center}
\begin{tabular}{ll}
        \hline
        \hline
	Level\_1 & at least one central calorimeter tower with $\iet\ > 5\ \gev$;\\
	 	& at least two tracks: one track with $\ippt\ > 5.48\ \pgev$, one with $\ippt\ > 2.46\ \pgev$\\
	Level\_2 & veto events with a calorimeter cluster with $\iet\ > 5\ \gev$, $1.1 < |\eta| < 3.6$;\\
	 	& require at least two clusters $\iet\ > 5\ \gev$, $|\eta| < 1.1$, which have $135 < \Delta \phi < 180$;\\
	 	& at least two SVT tracks with $\ippt\ > 2\ \pgev$, $160\ \mu \mbox{m} < |d_{0}| < 1000\ \mu \mbox{m}$, $\chi^{2} < 12$\\
	(OS)    & tracks have $150^{\circ} < \Delta \phi < 180^{\circ}$\\
	(SS)    & tracks have $2^{\circ} < \Delta \phi < 30^{\circ}$\\
	Level\_3 & at least two $\Delta R = 0.7$ jets with $\iet\ > 10\ \gev$, $|\eta| < 1.1$;\\
		& at least two SVT tracks with $\ippt\ > 2\ \pgev$, $|\eta| < 1.2$, $160\ \mu \mbox{m} < |d_{0}| < 1000\ \mu \mbox{m}$;\\
	 	& at least two COT tracks with $\ippt\ > 1.5\ \pgev$ $|\eta| < 1.2$, $130\ \mu \mbox{m} < |d_{0}| < 1000\ \mu \mbox{m}$,\\ 
	 	& track impact parameter significance $S(d_{0}) > 3$, $|\Delta z| < 5\ \mbox{cm}$\\
        \hline
	\hline
\end{tabular}
\end{center}
\end{table*}

At level\_2, there are two different paths, called opposite side (OS)
and same side (SS), which refer to the topological configuration of
the displaced tracks in the event. Both paths contain a veto on jets
in the plug calorimeter and a central calorimeter requirement.  The
plug jet veto requires that there are no calorimeter clusters with
$\iet\ > 5\ \gev$ in $|\eta| > 1.1$ and is designed to reduce the QCD
background, which produces more gluon radiation than does the signal.
The trigger requires at least two central calorimeter clusters, $\iet\
> 5\ \gev$ and $|\eta| < 1.1$, which are on opposite sides of the
calorimeter.  Finally, the calorimeter clusters must have in total at
least two displaced SVT tracks, with track $\ippt\ > 2\ \pgev$, $160\
\mu \mbox{m} < |d_{0}| < 1000\ \mu \mbox{m}$, secondary vertex fit
$\chi^{2} < 12$. For the OS path, the two tracks must have
$150^{\circ} < \Delta \phi < 180^{\circ}$.  The SS path requires that
the two displaced tracks point to a single cluster, and that
$2^{\circ} < \Delta \phi < 30^{\circ}$.

At level\_3, the trigger requires at least two jets with $\iet\ > 10\
\gev$ and $|\eta| < 1.1$. Jet clustering uses a cone algorithm of size
$\Delta R = 0.7$. The trigger also requires at least two tracks with
$160\ \mu \mbox{m} < |d_{0}| < 1000\ \mu \mbox{m}$, to confirm the
level\_2 requirements. As a cross-check this requirement is performed
with both SVT tracks and COT tracks, with additional impact parameter
significance and $|\Delta z|$ requirements imposed on the COT
tracks. The track parameter $z_{0}$ is the distance in the $z$
direction from the detector origin to the point on the z axis closest
to the track helix. The quantity $|\Delta z|$ is the magnitude of the
difference in $z_{0}$ between the two COT tracks. The cut on $|\Delta
z|$ ensures both tracks originate from the same primary vertex. The
same level\_3 requirements are imposed on both OS and SS level\_2 paths.

After the trigger selection, there is a further jet
classification. Jets in the analysis stage are reconstructed with a
$\Delta R = 0.4$ cone. The larger radius jet cone used in the trigger
provides high trigger efficiency. In the analysis, a cone of $0.4$ is
used to avoid unnecessary merging of jets. The \iet\ of the jet, after
being corrected to the hadron scale ($E_{T}^{\mathrm{cor}}$) must be greater
than $20\ \gev$. The jets must be in the central region of the
detector, $|\eta| < 1.0$. This requirement overlaps with the ZBB
trigger requirement. These jets are referred to as ``tight-central''
jets. However, as explained below, there are instances where
non-tight-central jets are used in this analysis.

\subsection{Signal and control regions}
\label{subsec:ES_signalcontrol}

While the HV phenomenology predicts four jets in the final state, we
allow events with three jets in order to increase our acceptance. In
addition, the plug jet veto in the ZBB trigger at level\_2 reduces jet
multiplicity (while simultaneously lowering the trigger rate at high
luminosity). Thus, in order to maintain acceptance, the signal region is
defined with three or more tight-central jets.

The signal MC samples show that the opening angle between the two jets
is not back-to-back, but instead usually smaller. The lighter the HV
particle, the smaller the opening angle between the jets. Thus, for each
jet pair in a 3-or-more jet event, we apply a cut of $\Delta R <
2.5$ on each pair. Events which pass the jet multiplicity cut and
have a jet pair passing the $\Delta R$ cut are said be in the signal
region.

In addition to the signal region, we define a two-jet control region to
validate our background estimation technique on a set of events that
is devoid of signal. This control region is defined as follows: events
are required to have exactly two tight-central jets with no $\Delta R$
requirement, and if additional jets are present, they must have
uncorrected $\iet\ < 15\ \gev$. The control and signal regions are
mutually exclusive.

\subsection{Secondary vertex tagging} 
\label{subsec:BE_secondaryvertex}

Secondary vertex tagging is used in this analysis to identify jets
with displaced vertices. We modified the standard CDF algorithm
{\sc SECVTX}~\cite{SECVTX} to increase the efficiency for very long-lived
particles, such as the HV particle, by extending the maximum impact
parameter allowed for tracks used in vertexing. Twenty $|d_{0}|_{max}$
cuts between $0.15\ \mbox{cm}$ and $1.6\ \mbox{cm}$ are studied to
maximize the signal while minimizing the increase in mistags. (Mistags
are light-flavor quark or gluon jets erroneously tagged as having a
displaced vertex.)


\section{Background Estimation}
\label{sect:backgroundestimation}

We want to produce a background estimate that retains the kinematic
correlations in QCD multijet events but has secondary vertices modeled
on those observed in SM processes. Toward this end, we use the jet
kinematics of our primary data sample, and to each event, we add
secondary vertices whose properties come from other data samples. The
vertices in these samples are characterized by probability-density
functions (PDFs) as a function of jet energy, flavor, and the number
of displaced tracks available to the ZBB trigger.

The ability to find particles with displaced vertices relies on the
reconstruction of secondary vertices. These secondary vertices can
come from multiple SM sources, which are listed in
Table~\ref{table_smprocesses}.

\begin{table}[htb]
\caption{\label{table_smprocesses} Standard model processes that can result in jets with reconstructed displaced vertices.}
\begin{center}
\begin{tabular}{ll}
	\hline
	\hline
	{\bf Background} &	 {\bf SM Production} \\
	\hline
	$b$-quarks         &	 QCD \bbbar, \ttbar, $W$/$Z$+jets, $WZ$/$ZZ$\\
	$c$-quarks         &	 QCD \ccbar, $W$/$Z$+jet, $WZ$/$ZZ$\\
	light-flavor (mistags) & QCD \qqbar\ \& $gg$, hadronic $\tau$s\\
	\hline
	\hline
\end{tabular}
\end{center}
\end{table}

\subsection{Building PDFs}
\label{subsec:BE_buildingpdfs}

The first step in modeling the background is to build the standard
model secondary vertex PDFs in tagged jets. The PDFs are constructed
from data events, when possible, where the signal is not expected to
be present, and in effect encapsulate SM secondary vertex information.

The pdf variables are defined in the plane transverse to the
beam line. We define the variables for the secondary vertex with
respect to the parent jet momentum vector, also called the jet
axis. First, define $\vec{L_{xy}}$ as the two-dimensional vector from
the primary vertex to the secondary vertex. There are two components
to this $\vec{L_{xy}}$ vector, one parallel to the jet axis, one
perpendicular. These two components are the first two pdf variables
and are named $u$ and $v$. These two variables define the position of
a secondary vertex with respect to a jet axis.

The two-dimensional $\Delta \phi$ angle between the jet axis and the
secondary vertex momentum vector is the third PDF variable, named
$\alpha$. This variable defines the direction of a secondary vertex
with respect to a jet axis in the plane transverse to the beam.

We find that correlations exist among all three variables. To preserve
these correlations, we store these PDFs in three-dimensional
histograms.

We split these PDFs into three main categories for different quark
flavors: $b$-quark, $c$-quark and light-flavor jets. These are further
split into different bins of $E^{\mathrm{cor}}_{T}$ and the number of SVT
tracks satisfying the ZBB-trigger SVT requirement. We separate jets
into bins of zero, one, and two or more SVT tracks, as the ZBB trigger
requires two displaced tracks in the event. We choose this binning
because the PDFs' shapes are different in each bin, due to the
dependence of secondary vertex production on these quantities.

Different data sources are used to construct the different quark
flavor PDFs. We use a muon trigger with a relatively low $\ippt$
requirement to build the $b$-quark PDFs. These data are rich in $B$
hadrons which decay semileptonically. To select events, we use a
two-jet selection where one jet is required to have a muon present
within its jet cone, and both jets are tagged while being well
separated in the detector ($\Delta \phi > 2.0$). The nonmuon jet,
called the away-jet, is the jet used to construct the $b$-quark PDFs.

For the light-flavor PDFs, we use the various CDF QCD jet triggers
which collect a large number of QCD multijet data events. These data
must have their heavy flavor contribution subtracted in order to
isolate the light-flavor events with secondary vertices; in effect,
these are mistag PDFs. The flavor composition of the jet triggers is
calculated using QCD MC templates of $b$-quark, $c$-quark, and
light-flavor secondary vertices. (In general, we refer to the
secondary vertex modeling derived from data as PDFs, while reserving
the word ``template'' to describe information obtained from MC samples.)
{\sc Pythia} QCD MC samples were generated with multiple momentum
thresholds. The two samples we used in this analysis are QCD to
\bbbar\ and generalized QCD with no final-state filtering. The former
is mostly used when comparing the signal MC in order to determine
discriminants, see Sec.~\ref{sect:hiddenvalley}. The latter is used
to construct the MC templates for the purposes of estimating the
background.

The vertex mass squared is the square of the sum of the four-momenta
of the tracks that form the secondary vertex, where the mass of the
track four-momentum is set to the mass of the pion. We determine the
flavor composition of the jet trigger data by fitting the QCD MC
templates of vertex mass to the data. Heavy flavor shapes ($b$- and
$c$-quark contributions) are subtracted from the jet trigger data,
using the QCD MC as the source of the shapes of the $b$- and $c$-quark
distributions for $u$, $v$, and $\alpha$.

In order to ensure that the $b$- and $c$-quark shapes for subtraction
accurately represent the data, we compare the $b$-quark PDFs from the
muon trigger to the QCD MC simulation where the jets are matched to a
$B$ hadron. The ratio of the means of each PDF variable is calculated
as the scale factor. The shapes of the distributions of these
variables are the same after applying these scale factors to the PDF
variables for $b$-quark jets from the QCD MC simulation. These scale
factors ($SF_{u} \sim 0.99$, $SF_{v} \sim 1.39$, $SF_{\alpha} \sim
1.04$) then are applied to the PDF variables for both $b$-quark and
$c$-quark jets from the QCD MC simulation when generating the
templates used in the subtraction procedure.



Finally, $c$-quark jets are not readily identifiable in real
data. Therefore we use QCD MC simulation in order to collect jets for
the $c$-quark PDFs. This is not a serious limitation because we find
that the ZBB data has a very small charm-quark component.

\begin{figure}[!ht]
\includegraphics[width=3.4in]{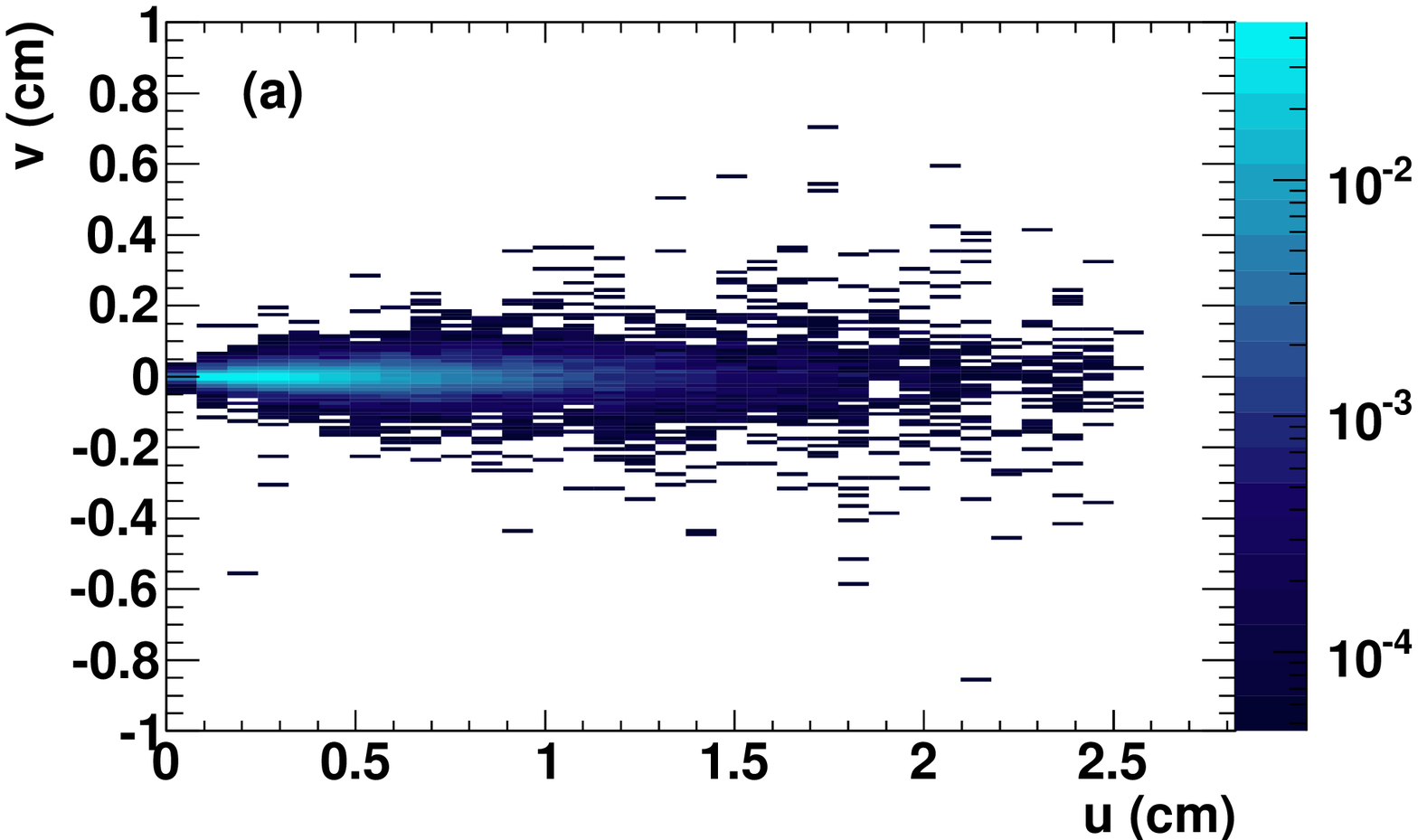}
\includegraphics[width=3.4in]{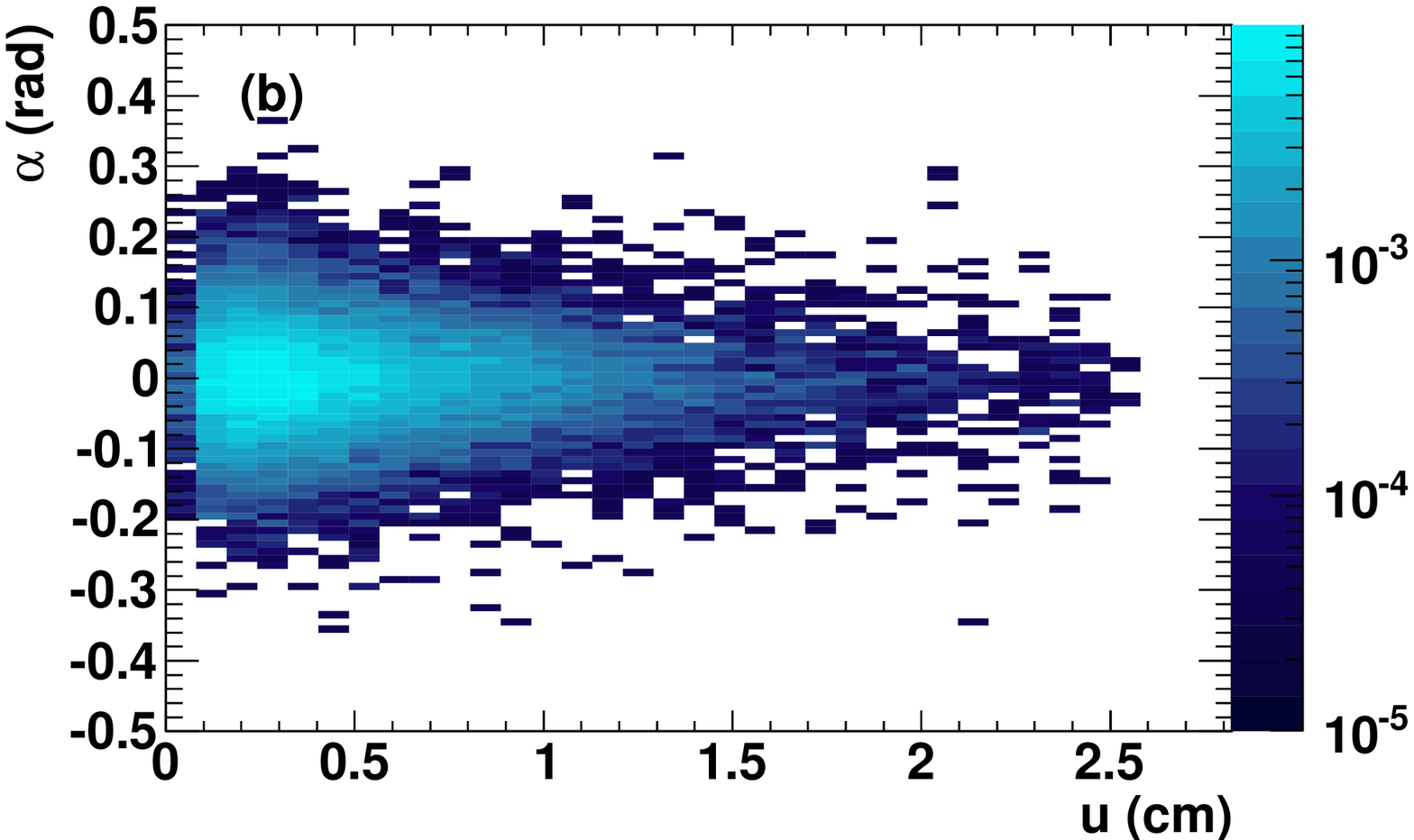}
\includegraphics[width=3.4in]{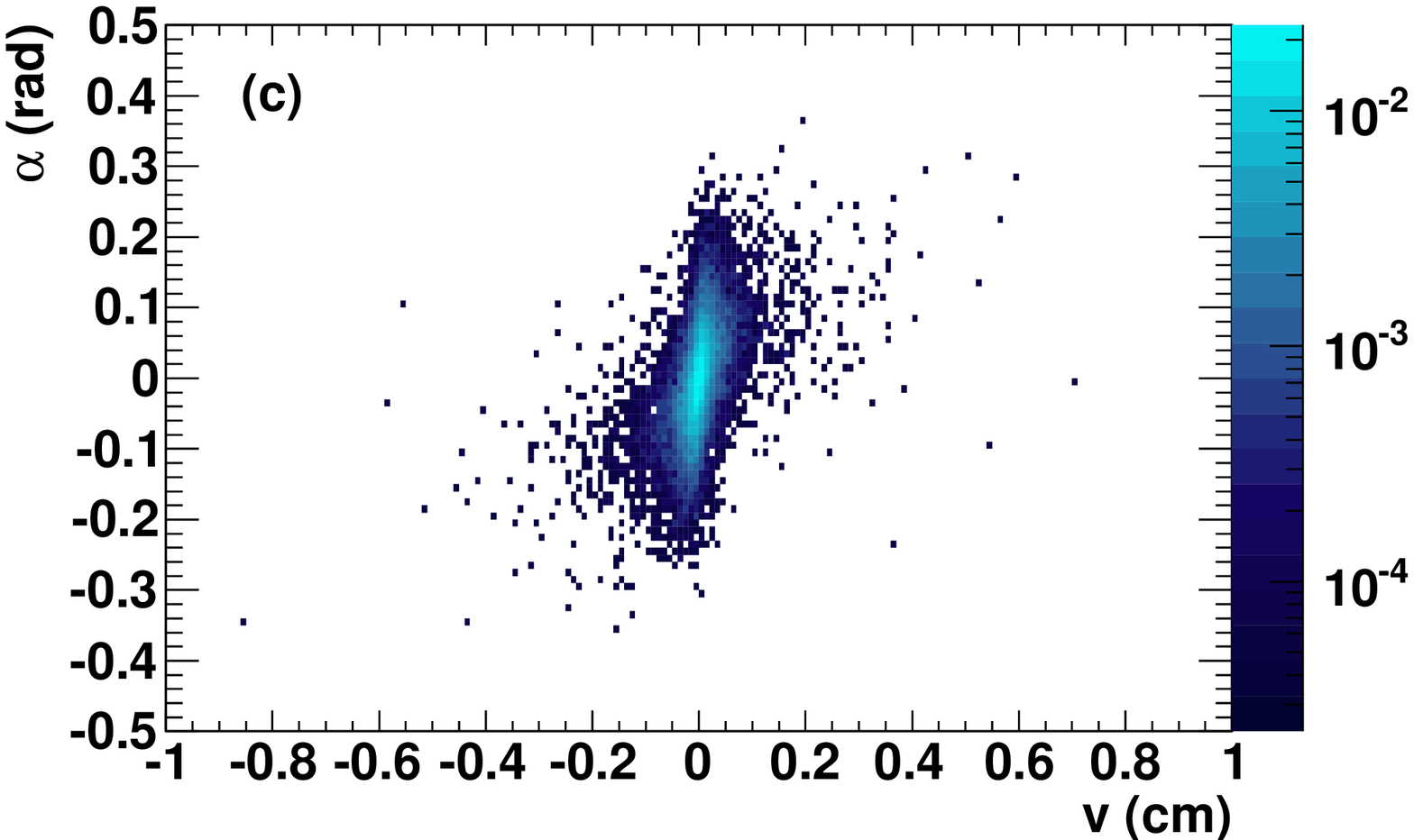}

\caption[Example PDFs]{
\label{fig_pdfexample}
Two-dimensional projections of a $b$-quark PDF for jets with \iet\
from $30$ to $70\ \gev$ and one SVT track within the jet: (a) $u$ vs
$v$, (b) $\alpha$ vs $u$, and (c) $\alpha$ vs $v$.}
\end{figure}

Two-dimensional projections of a $b$-quark PDF for jets with \iet\
from $30$ to $70\ \gev$ and one SVT track within the jet are shown in
Fig.~\ref{fig_pdfexample}.

\subsection{Pseudoevent generation} 
\label{subsec:BE_pseudoevent}

We build a background estimate using pseudoevents which are produced
by applying the secondary vertex PDFs to jets in events from the ZBB
trigger data. While the kinematic information comes from the real
event, the PDFs are used to characterize secondary vertices from SM
sources in the jets.

The background estimate is done using the same ZBB trigger sample used
to search for the signal. Events in the signal and control regions are
separately selected. When generating pseudoevents, a pair of
tight-central jets is used. For the signal region, this is the dijet
system where $\Delta R < 2.5$. (In principle, there could be multiple
dijets in a three-or-more-jet event, but in practice, we find only one
dijet system in each event.) In the control region, the dijet system
is simply the two tight-central jets in the event.

Before the pseudoevents are generated, we must understand the ZBB
trigger data. First, we obtain the dijet tag probability of real dijets
in the ZBB trigger data. This is the probability that {\bf both} of
the jets are tagged. The purpose of this dijet probability is to
preserve kinematic correlations that may exist with respect to
tagging.  The dijet probability is calculated in terms of both the
\iet\ of the jets and the number of SVT tracks, as the probability of a
tag changes with these variables.

Second, we obtain the flavor composition of the dijets in the ZBB
trigger data. However, unlike the flavor composition of the JET
trigger samples, where we were concerned with single jets, here we are
concerned with the flavor composition of pairs of jets. With three
possible flavor categories: $b$-quark (B), $c$-quark (C), and
light-flavor (L), and two jets, there are nine possible combinations of
double flavors for a pseudodijet: BB, BC, BL, CB, CC, CL, LB, LC, and
LL; where mixed states such as the BC and CB states are not considered
degenerate. The first letter describes the flavor for the leading
\iet\ jet and the second letter that of the subleading \iet\ jet.

\begin{figure}[htb]
\includegraphics[width=3.4in]{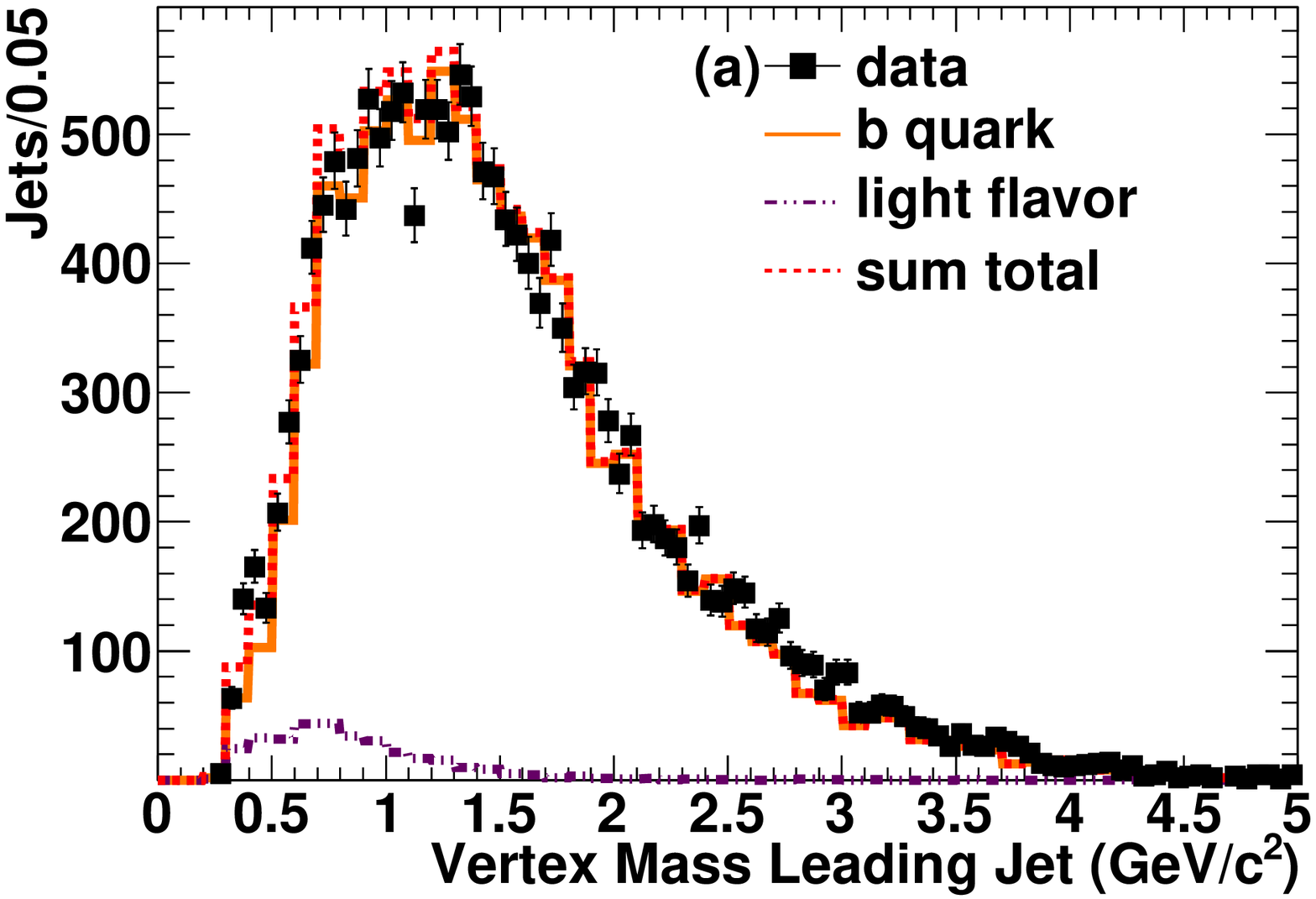}
\includegraphics[width=3.4in]{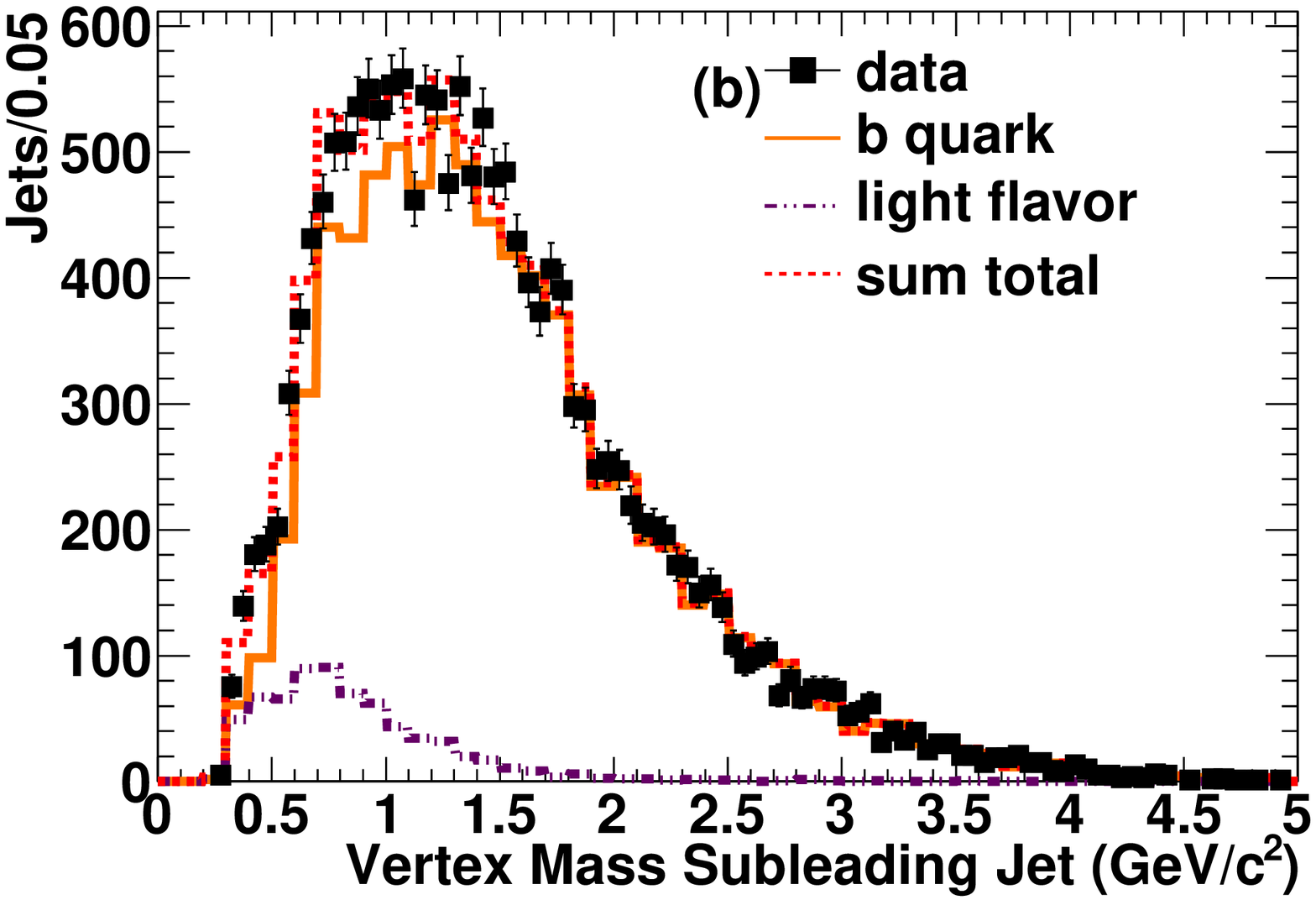}

\caption[Double-flavor fit]{
\label{fig_pedoubleflavorfit}
Example of double-flavor fits for dijets where each jet has exactly
one SVT trigger track. Histograms are projections of the
two-dimensional fit onto the axis of each jet: (a) the higher \iet\ jet
in the event, (b) the lower \iet\ jet.}
\end{figure}

We use two-dimensional fits of the vertex mass of the two vertices to
determine the flavor of jet pairs with tags. We reuse the {\sc Pythia}
QCD dijet MC templates of the vertex mass. First, the individual
$b$-quark, $c$-quark, and light-flavor MC templates are joined to form
two-dimensional vertex mass templates for BB, BC, BL, etc. Then, the
two-dimensional vertex mass templates are merged to form a single
vertex mass template that encompasses all nine double-flavor
states. Because the nine double flavor fractions must add to one,
there are eight fractions that we fit, which are algebraic
combinations of the nine double-flavor states. Fits are performed
using the {\sc RooFit} package~\cite{RooFit}; an example is shown
in Fig.~\ref{fig_pedoubleflavorfit}, and the fit result is shown in
Table~\ref{table_pedoubleflavor}.

\begin{table}[htb]
\caption{\label{table_pedoubleflavor}
Double-flavor fraction fit results for dijets where each jet has exactly
one SVT trigger track.}
\begin{center}
\begin{tabular}{cdcd}
        \hline
	\hline
	{\bf Double-Flavor State}       & \multicolumn{3}{c}{{\bf (\%)}} \\
        \hline
	BB              & 91.28 & $\pm$ & 0.96\\
	BC              & 0     & $\pm$ & 0.33\\
	BL              & 4.78  & $\pm$ & 0.91\\
	CB              & 0     & $\pm$ & 0.33\\
	CC              & 0     & $\pm$ & 0.17\\
	CL              & 0     & $\pm$ & 0.22\\
	LB              & 0.65  & $\pm$ & 0.53\\
	LC              & 0     & $\pm$ & 0.14\\
	LL              & 3.29  & $\pm$ & 0.82\\
	\hline
	\hline
\end{tabular}
\end{center}
\end{table}

The pseudoevent generation process is as follows: with a jet pair
selected, from either the control region or signal region, we proceed
to generate tags for the jets in the pair. The probability of both
jets having tags, calculated from the ZBB data, is used to assign
whether or not both jets have a tag in the pseudoevent.

Next, the flavor of the pseudojet is generated. Using
Table~\ref{table_pedoubleflavor} as an example, if both jets have
exactly one SVT trigger track, there is a 91.28\% probability that the
jet pair is BB, or two $b$-quark jets, a 4.78\% chance that the
pair is BL, where the leading \iet\ jet is a $b$-quark jet and the
subleading \iet\ jet is light-flavor, etc.

The secondary vertex information is sampled from the PDFs,
generated using background processes. In this step the jets are
sampled independently. The sampling is performed on the
three-dimensional histogram where the PDF information is stored.
Random $u$, $v$, and $\alpha$ are chosen according to the PDF's
distribution and assigned to the pseudojet.

To complete the process, a pseudoevent is generated for each event in
the ZBB trigger data (which are part of the control or signal
regions), thereby creating ``pseudodata'' with the proportion of
secondary vertices and the flavor composition derived from the ZBB
trigger data, and the secondary vertex information corresponding to SM
sources via the PDFs.

\subsection{Validation} 
\label{subsec:BE_validation}

\begin{figure}[htb]
\includegraphics[width=3.4in]{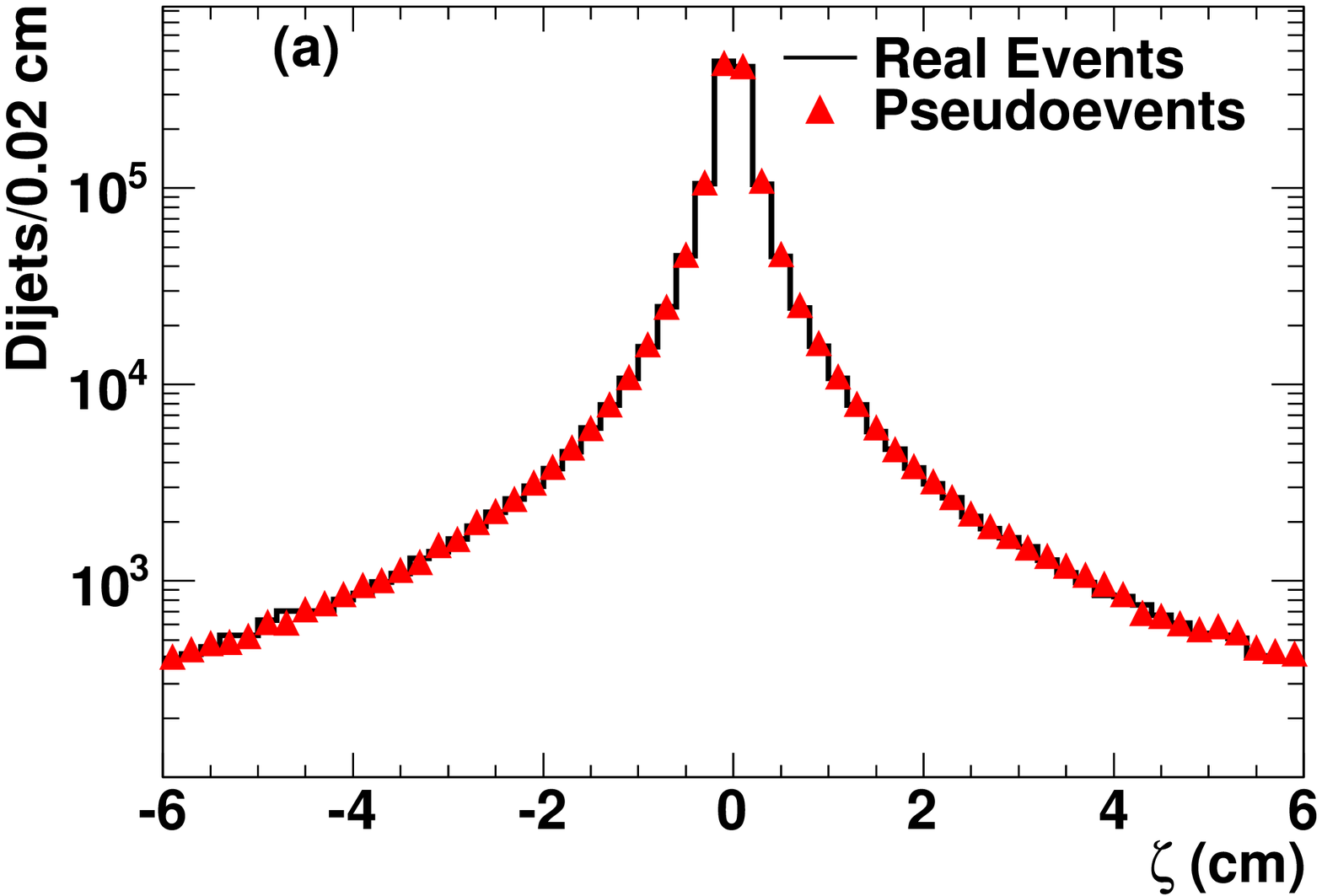}
\includegraphics[width=3.4in]{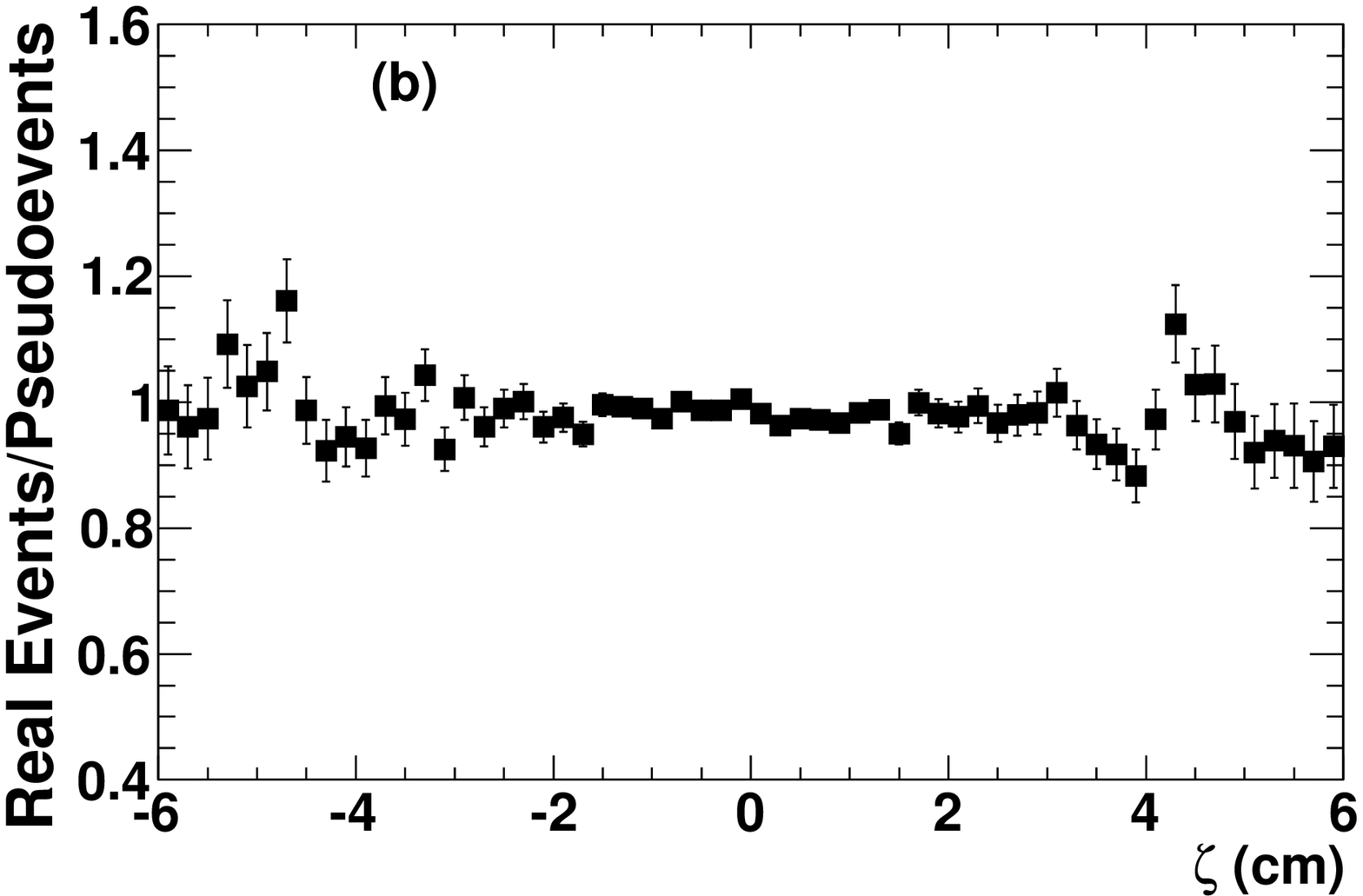}
\caption{\label{fig_controlzeta} The distribution of $\zeta$ for
the real events (solid line) and pseudoevents (triangle points) in
the control region (a), and their ratio (b).}
\end{figure}

We use the control region to validate this algorithm. Because we
expect the control region to be devoid of signal, we can compare the
real dijet data to the pseudodijets generated to see if the
pseudoevent generation replicates the data. For the purposes of this
validation, exactly one pseudoevent is generated for each real event,
and the PDFs are only sampled once for each pseudojet. Distributions
of the control region pseudoevent vs real events show that the
pseudoevent generation is well behaved. Figure~\ref{fig_controlzeta}
shows the distribution of $\zeta$ in the control region for real
events and pseudoevents, along with the ratio of the two
distributions.

\section{Signal Search}
\label{sect:signalsearch}

\subsection{From pseudoevents to a background estimate} 
\label{subsec:SS_backgroundestimate}

To generate the background estimate, we create pseudodata as described
above for events in the signal region. We construct a
``pseudoexperiment'' by sampling from the PDFs for each dijet in the
ZBB trigger sample. We carry out this procedure 10~000 times to create
an ensemble of pseudoexperiments. Each is treated independently and is
passed through the same set of analysis cuts, which will be described
in further detail. The resulting number of events that pass these cuts
is calculated for each pseudoexperiment. The background estimate is
the mean number of events that pass these cuts averaged over all
pseudoexperiments, and represents the number of events in ZBB trigger
data that would pass the analysis cuts if only SM processes
contributed to the observed data. We perform a simple counting
experiment by comparing this background estimate to the number of
observed data events with the same analysis cuts.

\subsection{Signal to background optimization} 
\label{subsec:SS_sigbackopt}

To conduct our search, we investigate the following variables,
optimizing the last three.

\newcounter{VARIABLES3}
\begin{list}{\arabic{VARIABLES3}.}{\usecounter{VARIABLES3}}
\item $|d_{0}|_{max}$ cut on tracks that are used by the tagging algorithm.
\item Separation of the two jets ($\Delta R$).
\item $\psi$, the impact parameter of a tagged jet.
\item $\zeta$, the decay distance of the HV particle.
\end{list}

\begin{figure}[htb]
\includegraphics[width=3.4in]{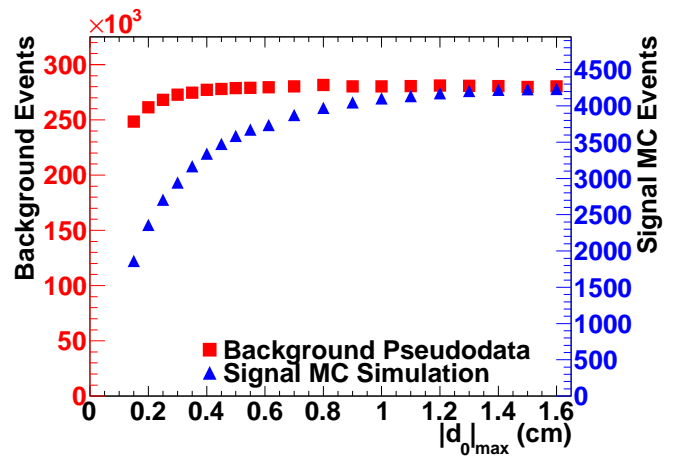}
\caption{\label{fig_SSsigbgdmaxd0} Number of signal MC simulation
events (triangle points) and background pseudoevents (square points)
vs the $b$-tagging algorithm cut $|d_{0}|_{max}$.}
\end{figure}

The $|d_{0}|_{max}$ cut is a parameter of the tagging
algorithm. Figure~\ref{fig_SSsigbgdmaxd0} shows the behavior of the
signal MC simulation and background pseudodata, where one pseudoevent
is generated for each real event, respectively, for the 20
$|d_{0}|_{max}$ cuts investigated. The standard CDF $|d_{0}|_{max}$
cut ($|d_{0}|_{max}<0.15\ \mbox{cm}$) reduces the efficiency of
finding secondary vertices from the signal MC events by more than
half.

The distribution for the signal shows that larger $|d_{0}|_{max}$ cuts
allow for more signal acceptance, while the background plateaus at
about $|d_{0}|_{max} = 0.70\ \mbox{cm}$. We choose the maximum
$|d_{0}|_{max}$ cut consistent with the physical constraints of the
CDF detector. The inner detector contains a beampipe with radius $r = 1.26\
\mbox{cm}$. Attached to the beampipe is a single layer of silicon
strips, Layer00 (L00). While it is not required that tracks deposit
hits in this inner most layer, we want to ensure that tracks
originating from the primary vertex {\em could} hit this
detector. Thus, the $d_{0}$ of tracks originating from a primary vertex
must be less than the radius of the beam pipe. However, the beam line
at CDF is not at the exact center of the detector. Accounting for this
shift, we determine that tracks with a maximum $|d_{0}| < 1.0\
\mbox{cm}$ may deposit hits in L00, thus a $|d_{0}|_{max} < 1.0\
\mbox{cm}$ is chosen.

With the $|d_{0}|_{max}$ cut set, we optimize the cuts on the other
three variables by maximizing $S/\sqrt{B}$.  The cuts chosen for the
low- and high-HV mass searches ($20\ \mgev$ for the former; $40$ and
$65\ \mgev$ for the latter) are shown in
Table~\ref{table_SSvarcuts}. The searches are optimized separately
because the low-mass HV results in daughter jets that are more
collinear. This changes the nature of the $\Delta R$ cut. For the
low-HV-mass search, only a $\Delta R_{max} < 0.75$ cut is imposed; no
$\Delta R_{min}$ cut is applied.

\begin{table}[htb]
\caption{\label{table_SSvarcuts}
Variable cuts for both the low- and high-HV mass searches.}
\begin{center}
\begin{tabular}{ccdccd}
        \hline
	\hline
	{\bf Variable}          & \multicolumn{2}{c}{\bf high HV} & \hspace{8mm} & \multicolumn{2}{c}{\bf low HV}\\
	                        & \multicolumn{2}{c}{\bf mass}    &               & \multicolumn{2}{c}{\bf mass}\\

        \hline
	$|d_{0}|_{max}$ (cm)    & $<$  & 1.0                      &               & $<$  & 1.0\\
	$\Delta R_{min}$        & $>$  & 0.75                     &               & \    & n.a.\\
	$\Delta R_{max}$        & $<$  & 2.0                      &               & $<$  & 0.75\\
	$|\psi|$ (both jets) (cm) & $>$  & 0.11                   &               & $>$  & 0.12\\
	$\zeta$ (cm)            & $>$  & 0.8                      &               & $>$  & 0.7\\
	$|\zeta|$ (cm)          & $<$  & \multicolumn{4}{c}{$\mbox{Minimum}(L_{xy\ 1}, L_{xy\ 2})$}\\
	\hline
	\hline
\end{tabular}
\end{center}
\end{table}

An additional cut shown in Table~\ref{table_SSvarcuts} is imposed
on $\zeta$. The magnitude of $\zeta$ must be less than the distance
from the primary vertex to the closest secondary vertex. This ensures
that the decay point is between the primary vertex and both
secondary vertices.

An unanticipated source of background became apparent when we applied
these analysis cuts to the real ZBB trigger events. A few events in
the low-HV mass sample appear to contain a single secondary vertex
from a $B$ hadron in which some of the decay products are found in
each of two nearby jet cones. This is a consequence of the small cone
size used in the jet algorithm. Since this is a physical background
that we had not thought of, we went to the signal MC to find criteria
that would remove this source of background but not adversely affect
the signal sensitivity. Two features of this background are that 1)
the two secondary vertices have very small separation in the
transverse plane ($\Delta S_{2d}$), and 2) the total invariant mass of
all the tracks in both vertices ($\Sigma m_{vtx}$) is less than the
$b$-quark mass. Table~\ref{table_SSvarcutsadd} shows the cuts made in
these variables. When these cuts are added to the low HV mass search,
the excess background described above is removed, while the reduction
in the efficiency in the signal MC simulation is negligible.

\begin{table}[htb]
\caption{\label{table_SSvarcutsadd}
Additional requirements on the low-HV-mass search due to an unanticipated background.}
\begin{center}
\begin{tabular}{ccd}
        \hline
        \hline
	{\bf Variable}          & \multicolumn{2}{c}{\bf low HV mass}\\
        \hline
	$\Delta S_{2d}$ (cm)    & $>$   & 0.06\\
        \multicolumn{3}{c}{OR}\\
	$\Sigma m_{vtx}$ ($\mgev$)   & $>$   & 5.0\\
	\hline
        \hline
\end{tabular}
\end{center}
\end{table}

\subsection{Results} 
\label{subsec:SS_results}

With the variable cuts set, we run 10~000 pseudoexperiments for both
mass searches to estimate the SM background. The distributions of the
number of pseudoevents passing the analysis cuts are Poisson
distributions with means $\mu_{low}=0.58$ and $\mu_{high}=0.29$. The
statistical uncertainties on these numbers are negligible. The
systematic uncertainty due to the background estimate procedure is
described in the next section.


With the same variable cuts, we can also calculate the number of
expected signal MC events that we would obtain with the same
integrated luminosity as the ZBB trigger. This is done by calculating
the number of events that pass the cuts in each signal MC sample and
multiplying this number by a scale factor consisting of the luminosity
of the ZBB trigger sample multiplied by the cross section for Higgs
boson production ($gg\rightarrow H$) divided by the number of signal
MC events generated. The Higgs boson cross sections are obtained from
Ref.~\cite{Junk:2010ar}. For $M_{H} = 130\ \mgev$, the cross section is
$\sigma_{gg\rightarrow H} = 858\ \mbox{fb}$, while for $M_{H} = 170\
\mgev$, the cross section is $\sigma_{gg\rightarrow H} = 349\
\mbox{fb}$. The branching ratio of the Higgs boson is assumed to be 100\% to
the HV particles, and the branching ratio of the HV particles is
assumed to be 100\% to \bbbar\ pairs.

When calculating the expected number of signal MC, two reweightings
are performed in order to account for differences between the ZBB
trigger data and signal MC events. First, we reweight to account for
differences in the luminosity profile of the signal MC events vs
data events. The distribution of the number of primary vertices in
data is divided by the same distribution in the MC events. The ratio
is used as an event-by-event weight, ranging from 0.75 to 5.0
depending on the number of primary vertices, in order to match the
luminosity profile of the signal MC simulation to the ZBB trigger
data.

A second reweighting is performed to account for different trigger
efficiencies for different run ranges in order to match the data's
trigger efficiency to that of the signal MC simulation.

To account for the differences between the MC and data tagging
efficiencies, a tagging scale factor for the tagging algorithm with
$|d_{0}|_{max}<1.0\ \mbox{cm}$ is also applied twice ($SF_{tagging}
= 0.9 \times 0.9 = 0.81$), because we have two tagged jets~\cite{CDF}.

Finally, a scale factor ($SF_{trigger} = 1.12 \pm 0.11$) is applied to account
for systematic effects present in the ZBB trigger simulation used on
the signal MC events~\cite{Donini2008nt}.

Table~\ref{table_SSresults} shows the results of our search. The
number of expected signal events is calculated from each of the signal
MC samples. The number of background events is also shown. Both
uncertainties are calculated in the next section. Of the
$6.2\ \times\ 10^{6}$ data events in the signal region, 124~000 of
which have two tagged jets, one event passes the analysis cuts in the
low-HV mass search, and one different event passes these cuts in the
high-HV mass search.

\begin{table*}[htb]
\caption{\label{table_SSresults} 
Results of the search. Two events remain after all cuts: one event in the low-HV-mass search, the second (different event) in the high-HV-mass search. Uncertainties are discussed in Sec.~\ref{sect:systematicuncertainties}.}
\begin{center}
\begin{tabular}{ccdddc}
        \hline
        \hline
        {\bf Higgs Boson}        & {\bf HV Mass} & \multicolumn{1}{c}{{\bf HV life-}}        & \multicolumn{1}{c}{{\bf Expected}}        & \multicolumn{1}{c}{{\bf Background}}      & {\bf Number}\\
	{\bf Mass ($\mgev$)}         & {\bf ($\mgev$)} & \multicolumn{1}{c}{{\bf time (cm)}}       & \multicolumn{1}{c}{{\bf Signal MC}}       & \multicolumn{1}{c}{{\bf Estimate}}        & {\bf Observed}\\
	\hline
	\multicolumn{6}{c}{low HV mass search}\\
	130                     & 20            & 1.0                   & 0.64                  & 0.58                 & 1\\
	170                     & 20            & 1.0                   & 0.074                 & 0.58                 & 1\\
	\\
	\hline
	\multicolumn{6}{c}{high HV mass search}\\
	130                     & 40            & 1.0                   & 0.26                  & 0.29                 & 1\\
	170                     & 40            & 1.0                   & 0.38                  & 0.29                 & 1\\
	170                     & 65            & 1.0                   & 0.14                  & 0.29                 & 1\\
	130                     & 40            & 0.3                   & 0.24                  & 0.29                 & 1\\
	130                     & 40            & 2.5                   & 0.10                  & 0.29                 & 1\\
	130                     & 40            & 5.0                   & 0.043                 & 0.29                 & 1\\
        \hline
	\hline
\end{tabular}
\end{center}
\end{table*}


\section{Systematic Uncertainties}
\label{sect:systematicuncertainties}

The systematic uncertainties in this analysis fall into two main
categories. The first are systematic effects that affect the
background estimate. The second are the systematic effects that affect
the number of expected signal MC events.

In the first category, there are three major sources of uncertainty,
each corresponding to a step in the pseudoevent generation. First,
the tagging probability is shifted by its statistical uncertainty one
sigma in each direction. The results are propagated through as a
systematic uncertainty.

The flavor composition probabilities used to determine the
pseudoflavor of the jets result in two more systematic uncertainties:
the statistical uncertainty from the flavor composition fractional fit
and a systematic due to the MC simulation over-efficiency in track
reconstruction, which has a direct impact on the vertex mass of the
secondary vertex. When additional tracks in the MC are reconstructed,
they will add to the vertex mass of the secondary vertex. For the
former, we use one sigma shifts in both directions. For the latter, we
use an overall 3\% reduction in the vertex mass to model a maximal
variation this over-efficiency would produce~\cite{gammametb}. This
reduction changes the flavor composition, and propagates through as a
systematic uncertainty.

We generate five new ensembles of pseudoevents where each is generated
with one of the systematic shifts described above. For each, we
perform another 10~000 pseudoexperiments as before and compare the
background estimate to the central value calculated in
Table~\ref{table_SSresults}. The percent difference is taken as the
systematic uncertainty.

For the last of the three steps in pseudoevent generation, the
PDF sampling, we turn to the bootstrap technique~\cite{Efron}. The
bootstrap technique measures the systematic uncertainty arising from
shape uncertainties in the PDFs. Effectively the three-dimensional
histograms are statistically varied within their Poisson statistical
fluctuations, and a new background estimate is calculated using 10~000
new pseudoexperiments. This procedure is itself replicated 200
times. The standard deviation of these 200 means is the uncertainty
associated with the PDF sampling.

The second category of systematic uncertainties affects the signal
estimate. These include uncertainties associated with:

\newcounter{SIGMCSYS4}
\begin{list}{\arabic{SIGMCSYS4}.}{\usecounter{SIGMCSYS4}}
\item jet energy scale,
\item trigger simulation,
\item tagging scale factor,
\item parton distribution function,
\item luminosity.
\end{list}

\noindent The first systematic uncertainty is calculated separately
for each signal MC sample, while the systematic uncertainties on the
trigger simulation, tagging scale factor, and luminosity have the same
value across all samples, and the parton distribution function has
approximately the same value for all samples.

The JES factor on jets is varied up (down) one sigma with respect to
its central value. The result is that more (less) jets pass the
$E_{T}^{\mathrm{cor}} > 20\ \gev$ cut. This affects the number of expected signal
MC events differently for each sample. 

The ZBB trigger simulation scale factor has an uncertainty of
8.9\%. The scale factor systematic uncertainty for the tagging
algorithm at the operating point, $|d_{0}|_{max} < 1.0\ \mbox{cm}$, is
10\%. The parton distribution function uncertainty is taken from
Ref.~\cite{Junk:2010ar} which documents this uncertainty for multiple
analyses, including ones that use $gg \rightarrow H$ production
(2.5\%). Finally, the luminosity uncertainty is 6\%~\cite{luminosity}.

All the systematic uncertainties calculated are shown in
Table~\ref{table_UNmain2}.  These systematic uncertainties are used in
the calculation of the limits discussed in
Sec.~\ref{sect:conclusion}.

\begin{table}[htb]
\caption{\label{table_UNmain2}
Summary of systematic uncertainties for the background estimate and
signal MC simulation. The JES is calculated separately for each signal
MC sample.}
\begin{center}
\begin{tabular}{cdd}
	\hline
	\hline
	{\bf Uncertainty}       & \multicolumn{1}{c}{{\bf Down (\%)}}    & \multicolumn{1}{c}{{\bf Up (\%)}}\\
	\hline
	\multicolumn{3}{c}{Background estimate - low HV mass search}\\
	Data statistics         & \multicolumn{2}{c}{$\pm$0.039}\\
	Tag prob. statistics    & -7.7  	& 3.4\\
	Flavor composition      & -0.5 	        & 2.75\\
	\\
	\hline
	\multicolumn{3}{c}{Background estimate - high HV mass search}\\
	Data statistics         & \multicolumn{2}{c}{$\pm$0.046}\\
	Tag probability statistics    & \multicolumn{2}{c}{$\pm$3.9}\\
	Flavor composition      & -0.5 	        & 8.9\\
	\\
	\hline
	\multicolumn{3}{c}{Signal MC}\\
	Jet Energy Scale        & \multicolumn{1}{c}{-15.6 to -6.3}  & \multicolumn{1}{c}{4.0 to 25.5}\\
	Trigger Uncertainty     & \multicolumn{2}{c}{$\pm$8.9}\\
	Tagging scale factor    & \multicolumn{2}{c}{$\pm$10}\\
	PDF                     & \multicolumn{2}{c}{$\pm$2.5}\\
	Luminosity              & \multicolumn{2}{c}{$\pm$6}\\
	\hline
	\hline
\end{tabular}
\end{center}
\end{table}


\section{Final Results and Conclusion}
\label{sect:conclusion}

With all the uncertainties calculated, we form test hypotheses
consisting of our background estimate along with our signal MC. A
separate test hypothesis is constructed for each set of masses and
lifetimes. We also create corresponding null hypotheses consisting
only of the background estimate for each HV mass
search. Table~\ref{table_CONpvalues} shows p-values for each set of
masses, showing the probability that the null hypothesis has
fluctuated to the data.

\begin{table}[htb]
\caption{\label{table_CONpvalues}
Null hypothesis p-values for this search.}
\begin{center}
\begin{tabular}{ccdd}
        \hline
        \hline
        {\bf Higgs Boson}        & {\bf HV Mass}   & \multicolumn{1}{c}{{\bf HV life-}}        & \multicolumn{1}{c}{{\bf p-value}}\\
	{\bf Mass ($\mgev$)}     & {\bf ($\mgev$)} & \multicolumn{1}{c}{{\bf time (cm)}}       & \\
	\hline
	\multicolumn{4}{c}{low HV mass search}\\
	130                     & 20            & 1.0 		        & 0.44\\
	170                     & 20            & 1.0 		        & 0.43\\
	\\
	\hline
	\multicolumn{4}{c}{high HV mass search}\\
	130                     & 40            & 1.0 		        & 0.27\\
	170                     & 40            & 1.0 		        & 0.26\\
	170                     & 65            & 1.0 		        & 0.26\\
	130                     & 40            & 0.3 		        & 0.27\\
	130                     & 40            & 2.5 		        & 0.27\\
	130                     & 40            & 5.0 		        & 0.27\\
	\hline
        \hline
\end{tabular}
\end{center}
\end{table}

We do not observe a statistically significant excess, thus we proceed
to set a limit on the production cross section times branching ratio
of the HV model for the particular masses and lifetimes we studied. A
Bayesian limit calculator is used for this
calculation~\cite{Junk:1999kv,PDG}. Table~\ref{table_CONlimits}
shows the resulting observed limit and median expected limit, along
with the $\pm1$ and $\pm2$ sigma values on the expected limit, all at
95\% confidence level (C.L.).

\begin{table*}[htb]
\caption{\label{table_CONlimits}
Observed and expected limits at 95\% C.L. calculated for different signal MC samples.}
\begin{center}
\begin{tabular}{ccddddddd}
        \hline
        \hline
        {\bf Higgs Boson}        & {\bf HV Mass}   & \multicolumn{1}{c}{{\bf HV life-}}        & \multicolumn{1}{c}{{\bf Observed}}      & \multicolumn{5}{c}{{\bf Expected Limit (pb)}}\\
	{\bf Mass ($\mgev$)}     & {\bf ($\mgev$)} & \multicolumn{1}{c}{{\bf time (cm)}}       & \multicolumn{1}{c}{{\bf Limit (pb)}}    & \multicolumn{1}{c}{-2 $\sigma$}   & \multicolumn{1}{c}{-1 $\sigma$}   & \multicolumn{1}{c}{median}        & \multicolumn{1}{c}{+1 $\sigma$}   & \multicolumn{1}{c}{+2 $\sigma$}\\
	\hline
	\multicolumn{9}{c}{low HV mass search}\\
	130                     & 20            & 1.0 		        & 6.2  	                & 4.3 	& 4.3 	& 4.3 	& 6.2  	& 8.4 \\
	170                     & 20            & 1.0 		        & 22.1                  & 15.2  & 15.2  & 15.2  & 22.1 	& 29.9 \\
	\\
	\hline
	\multicolumn{9}{c}{high HV mass search}\\
	130                     & 40            & 1.0 		        & 15.9 	                & 10.5	& 10.5 	& 10.5 	& 15.9 	& 21.5 \\
	170                     & 40            & 1.0 		        & 4.4  	                & 2.9  	& 2.9  	& 2.9  	& 4.4  	& 6.0 \\
	170                     & 65            & 1.0 		        & 11.7 	                & 7.7  	& 7.7  	& 7.7  	& 11.7 	& 15.7\\
	130                     & 40            & 0.3 		        & 17.8 	                & 11.7 	& 11.7 	& 11.7 	& 17.8 	& 24.2 \\
	130                     & 40            & 2.5 		        & 40.7 	                & 26.8  & 26.8 	& 26.8 	& 40.7 	& 55.1 \\
	130                     & 40            & 5.0 		        & 94.3 	                & 62.0  & 62.0 	& 62.0 	& 94.3 	& 127.9\\
	\hline
        \hline
\end{tabular}
\end{center}
\end{table*}

The counting experiment is performed with a small discrete number of
events, where the background estimate is less than one. Thus, the
expected number of events can only fluctuate up (from zero). The
result is that the negative sigma expected limits are identical to the
median limit. Also, the $+1$ sigma expectation is 1 event, which is
what we see. Thus, our limit is the same as the $+1$ sigma expectation.

\begin{figure}[htb]
\includegraphics[width=3.4in]{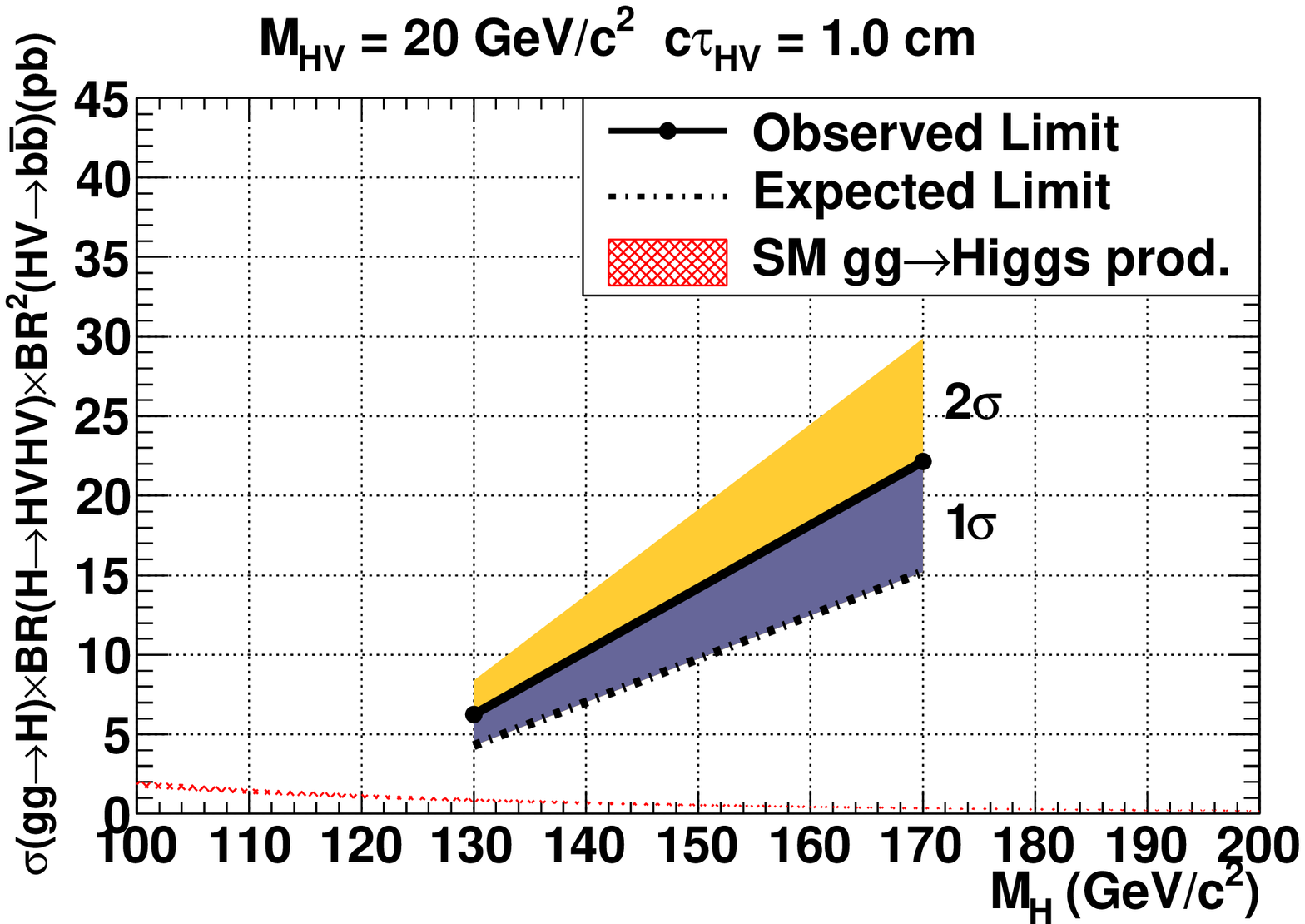}
\caption{\label{fig_CONlimit1} 
Observed and expected limits at 95\% C.L. with $+1$ and $+2\ \sigma$
bands for signal MC simulation with HV mass $20\ \mgev$. The hashed
line is the Higgs boson production cross section in the SM.}
\end{figure}

\begin{figure}[htb]
\includegraphics[width=3.4in]{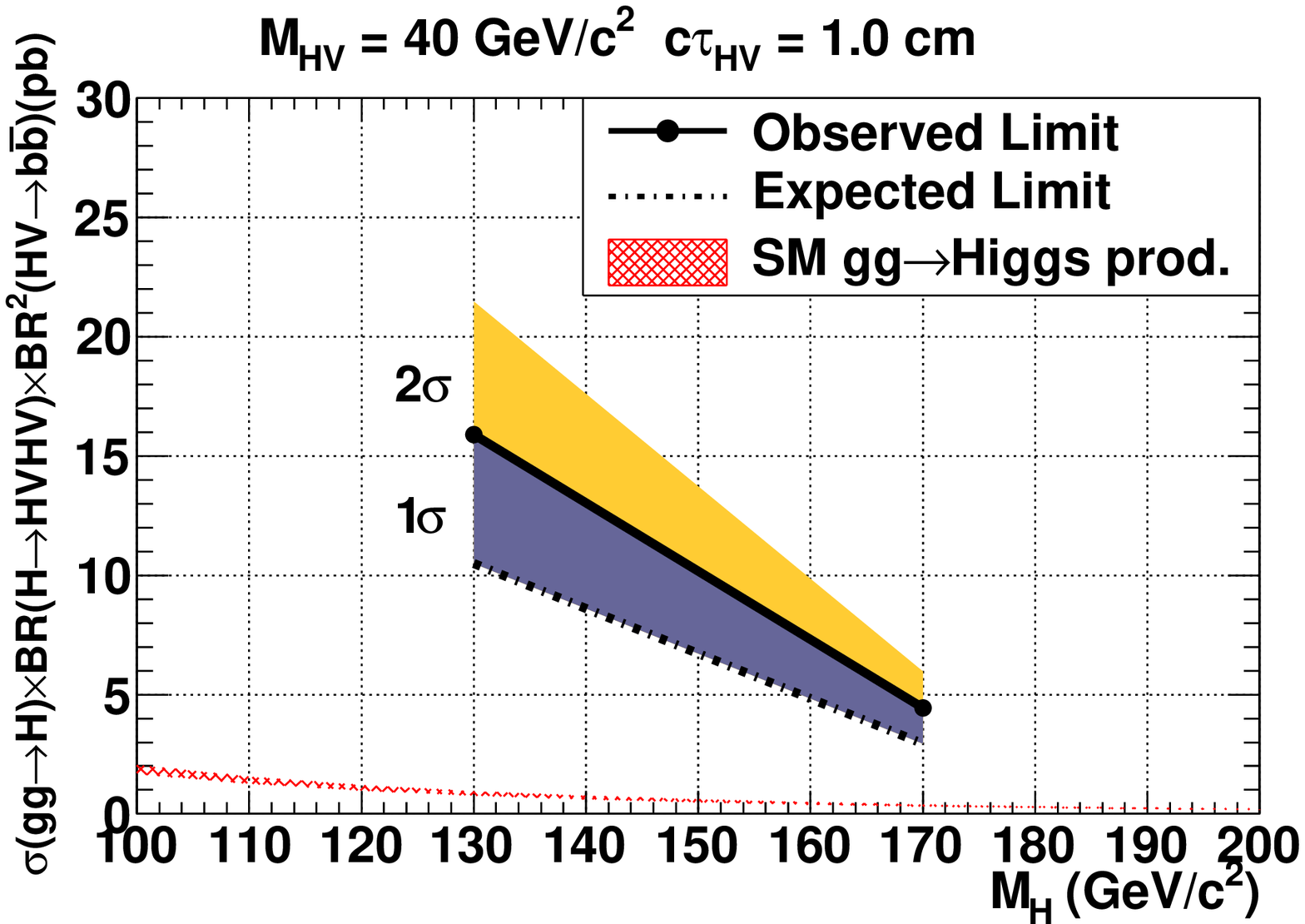}
\caption{
\label{fig_CONlimit2}
Observed and expected limits at 95\% C.L. with $+1$ and $+2\ \sigma$
bands for signal MC with HV masses $40\ \mgev$. The hashed line is the
Higgs boson production cross section in the SM.}
\end{figure}

\begin{figure}[htb]
\includegraphics[width=3.4in]{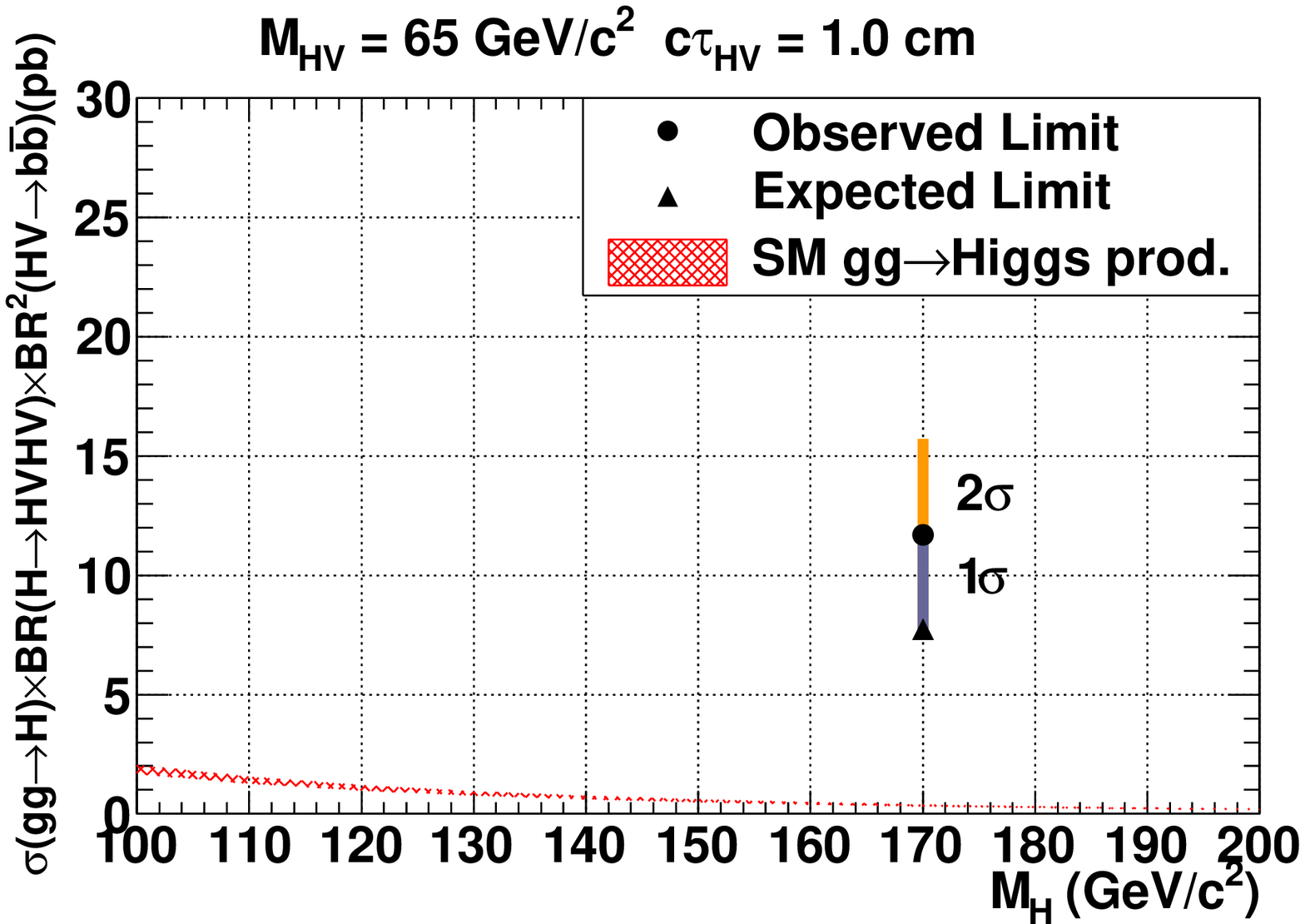}
\caption{\label{fig_CONlimit3}
Observed and expected limits at 95\% C.L. with $+1$ and $+2\ \sigma$
bands for signal MC with HV masses $65\ \mgev$. The hashed line is the
Higgs boson production cross section in the SM.}
\end{figure}

\begin{figure}[htb]
\includegraphics[width=3.4in]{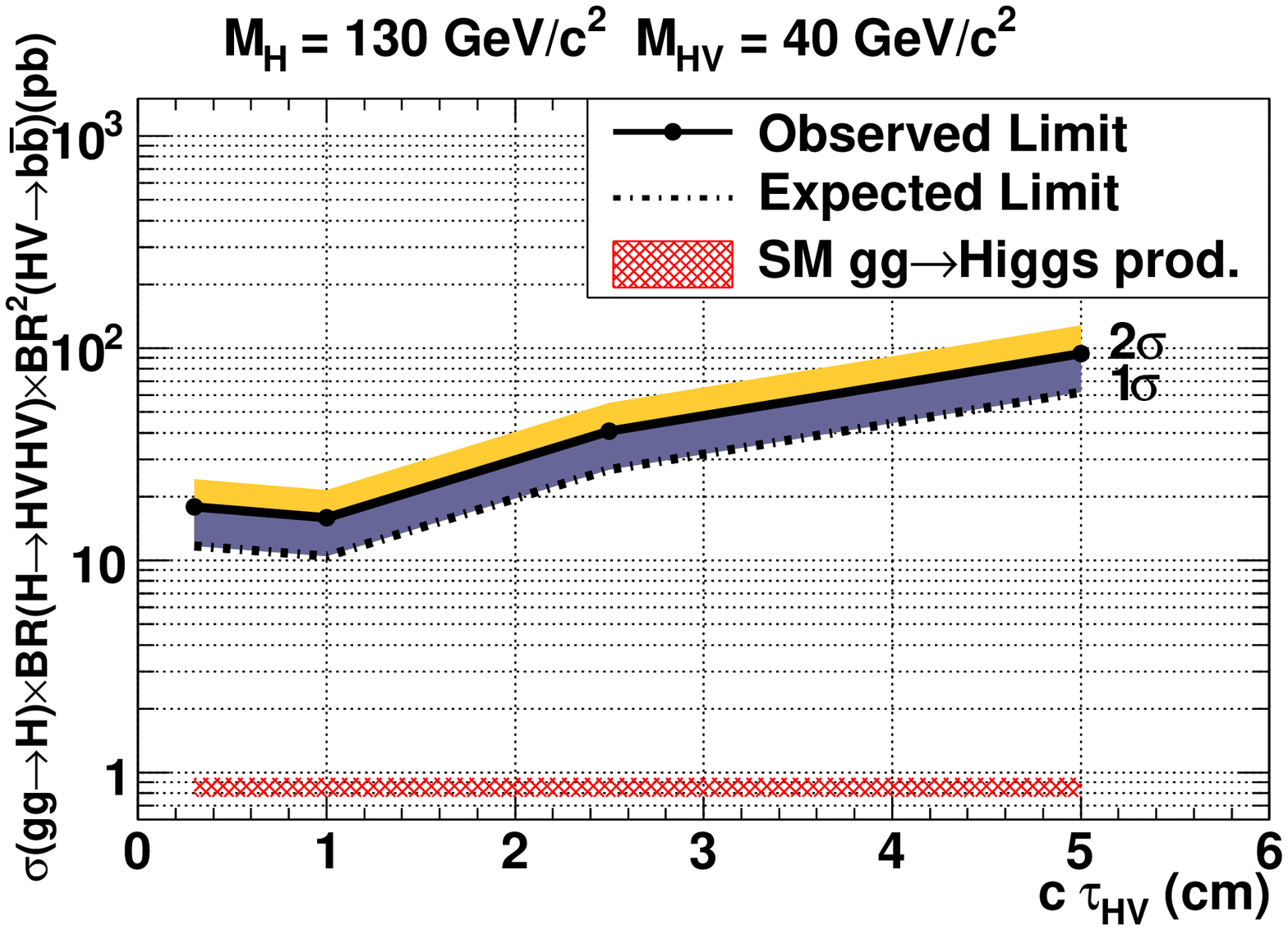}
\caption{
\label{fig_CONlimit4}
Observed and expected limits at 95\% C.L with $+1$ and $2\ \sigma$
bands for signal MC for differing HV particle lifetimes. The hashed
line is the Higgs boson production cross section in the SM.}
\end{figure}

Figures~\ref{fig_CONlimit1},~\ref{fig_CONlimit2},~\ref{fig_CONlimit3}
and~\ref{fig_CONlimit4} show the results of the limit calculation. In
Figs.~\ref{fig_CONlimit1},~\ref{fig_CONlimit2}
and~\ref{fig_CONlimit3}, the x-axis is the mass of the Higgs
boson. Figure~\ref{fig_CONlimit1} is for a $M_{\mathrm{HV}}$ of $20\ \mgev$
corresponding to the low-HV-mass search. Figure~\ref{fig_CONlimit2} is
for a $M_{\mathrm{HV}}$ of $40\ \mgev$, corresponding to the high-HV-mass
search. Figure~\ref{fig_CONlimit3} shows the results of the
high-HV-mass search for a $M_{\mathrm{HV}}$ of
$65\ \mgev$. Figure~\ref{fig_CONlimit4} shows the limits for $M_{H}$
of $130\ \mgev$ and $M_{\mathrm{HV}}$ of $40\ \mgev$ with the HV lifetime on
the x-axis.

In conclusion, we have searched for heavy metastable particles that
decay into a jet pair at a displaced vertex at CDF. No statistically
significant excess is observed, and limits are set on the production
cross section times branching ratio for the HV phenomenology we have
used as a benchmark. The results shown for this phenomenology can be
used to constrain other models by considering the differences of the
cross section, branching ratio, and the kinematics of the final state.

\clearpage


\begin{acknowledgments}
We thank the Fermilab staff and the technical staffs of the
participating institutions for their vital contributions. This work
was supported by the U.S. Department of Energy and National Science
Foundation; the Italian Istituto Nazionale di Fisica Nucleare; the
Ministry of Education, Culture, Sports, Science and Technology of
Japan; the Natural Sciences and Engineering Research Council of
Canada; the National Science Council of the Republic of China; the
Swiss National Science Foundation; the A.P. Sloan Foundation; the
Bundesministerium f\"ur Bildung und Forschung, Germany; the World
Class University Program, the National Research Foundation of Korea;
the Science and Technology Facilities Council and the Royal Society,
UK; the Institut National de Physique Nucleaire et Physique des
Particules/CNRS; the Russian Foundation for Basic Research; the
Ministerio de Ciencia e Innovaci\'{o}n, and Programa
Consolider-Ingenio 2010, Spain; the Slovak R\&D Agency; and the
Academy of Finland.
\end{acknowledgments}



\begin{thebibliography}{99}
\bibitem{Strassler2007374}
  M.~Strassler and K. Zurek, Physics Letters B {\bf 651}, 374 (2007). 

\bibitem{PhysRevLett.103.071801}
  V.~M.~Abazov, {\it et al}. Int.\ J.\ Mod.\ Phys., A {\bf 20}, 3263 (2005).

\bibitem{Bardi:1997hm}
  A.~Bardi, {\it et al.}, Nucl.\ Instrum.\ Methods\ A {\bf 409}, 658 (1998).

\bibitem{Adelman:2007zz}
  J.~A.~Adelman {\it et al.} (CDF Collaboration), Nucl.\ Instrum.\ Methods\ A {\bf 572}, 361 (2007).

\bibitem{CDF}
  F.~Abe {\it et al.}, Nucl.\ Instrum.\ Methods {\bf 271}, 387 (1988);
  D.~E.~Acosta {\it et al.}  (CDF Collaboration), Phys.\ Rev.\  D {\bf 71}, 052003 (2005);
  The CDF-II Detector Technical Design Report, Fermilab-Pub-96/390-E.

\bibitem{SVX}
  A.~Sill {\it et al.}, Nucl.\ Instrum.\ Methods A {\bf 447}, 1 (2000);
  A.~Affolder {\it et al.}, Nucl.\ Instrum.\ Methods A {\bf 453}, 84 (2000);
  C.~S.~Hill {\it et al.},  Nucl.\ Instrum.\ Methods A {\bf 530}, 1 (2000).

\bibitem{COT}
  A.~Affolder {\it et al.},  Nucl.\ Instrum.\ Methods A {\bf 526}, 249 (2004).

\bibitem{coord}
The CDF~II uses a cylindrical coordinate system in which $\phi$ is the
azimuthal angle, $r$ is the radius from the nominal beam line, and $z$
points in the proton beam direction. The transverse plane, $r-\phi$
plane, or sometimes referred to as $x$-$y$ plane, is the plane
perpendicular to the $z$ axis. In addition, $\theta$ is defined as the
polar angle measured with respect to the interaction vertex.
Transverse momentum and energy are the projections of total momentum
and energy onto the $r-\phi$ plane and defined as $\ippt =
p\sin\theta$ and $\iet = E\sin\theta$, respectively.

\bibitem{CEM}
  L.~Balka {\it et al.}, Nucl.\ Instrum.\ Methods A {\bf 267}, 272 (1988);
  S.~R.~Hahn {\it et al.}, Nucl.\ Instrum.\ Methods  A {\bf 267}, 351 (1988).

\bibitem{CHA}
  S.~Bertolucci {\it et al.}, Nucl.\ Instrum.\ Methods, A {\bf 267}, 301 (1988);
  M.~Albrow {\it et al.}, Nucl.\ Instrum.\ Methods, A {\bf 480}, 524 (2002);
  G.~Apollinari {\it et al.}, Nucl.\ Instrum.\ Methods, A {\bf 412}, 515 (1998).

\bibitem{cal_upgrade} 
  S.~Kuhlmann {\it et al.}, Nucl.\ Instrum.\ Methods A {\bf 518}, 39, 2004.

\bibitem{jetclu}
  F.~Abe {\it et al.} (CDF Collaboration), Phys.\ Rev.\ D {\bf 45}, 1448 (1992).


\bibitem{CLC}
  D.~Acosta {\it et al.}, Nucl.\ Instrum.\ Methods A {\bf 494}, 57 (2002).

  M.~Strassler and K. Zurek, Physics Letters B {\bf 651}, 374 (2007). 

\bibitem{Strassler2008263} 
  M.~Strassler and K. Zurek, Physics Letters B {\bf 661}, 263 (2008). 

\bibitem{kaplanlutyzurek}
  D.~Kaplan, M.~Luty, and K.~Zurek, Phys.\ Rev.\ D {\bf 79}, 115016 (2009). 

\bibitem{pythia}
  T.~Sjöstrand, P.~Edén, C.~Friberg, L.~Lönnblad, G.~Miu, S.~Mrenna and E.~ Norrbin, Comput.~Phys.~Commun. {\bf 135} 238 (2001). The version of {\sc pythia} used here is 6.216 (arXiv:hep-ph/0108264).


\bibitem{GEANT3}
E.~Gerchtein and M.~Paulini, in {\em Proceedings of 2003 Conference for Computing in High-Energy and Nuclear Physics (CHEP 03), La Jolla, California}, 24-28 Mar 2003, pp TUMT005. The version of {\sc Geant} used here is 3.21/14, see CERN Program Library Long Writeup W5013.



\bibitem{Donini2008nt}
  J.~Donini {\it et al.}, Nucl.\ Instrum.\ Methods A {\bf 596}, 354 (2008). 

\bibitem{Bhatti2005ai}
  A.~Bhatti {\it et al.}, Nucl.\ Instrum.\ Methods A {\bf 566}, 375 (2006). 


\bibitem{SECVTX}
  D.~E.~Acosta {\it et al.} (CDF Collaboration),  Phys.\ Rev.\  D {\bf 71}, 052003 (2005);
  C.~Neu, FERMILAB-CONF-06-162-E.

\bibitem{RooFit} 
  W.~Verkerke and D.~Kirkby (unpublished).  For documentation and source code, see {\it http://roofit.sourceforge.net} and arXiv:physics/0306116.

\bibitem{Junk:2010ar}
  The TEVNPH Working Group (CDF and D0 Collaborations) (2010), arXiv:hep-ex/1007.4587.

\bibitem{gammametb}
  T.~ Aaltonen {\it et al.} (CDF Collaboration), Phys.\ Rev.\ D {\bf 80}, 052003 (2009).

\bibitem{Efron} 
  B.~Efron and R.~J.~Tibshirani, ``An Introduction to the Bootstrap''. (New York : Chapman \& Hall, 1993).

\bibitem{luminosity}
  D.~Acosta {\it et al.} (CDF Collaboration), Phys.\ Rev.\ Lett.\ 94, 091803 (2005).

\bibitem{Junk:1999kv}
  T.~Junk, Nucl.\ Instrum.\ Methods\ {\bf A434}, 435-443 (1999). 

\bibitem{PDG}
  K.~Nakamura {\it et al.} (Particle Data Group), J.\ Phys.\ G {\bf 37}, 075021 (2010).
\end{thebibliography}
\end{document}